\documentclass[aps,pre,preprint,showpacs,showkeys,floatfix,superscriptaddress]{revtex4-1}
\usepackage{graphicx}
\usepackage{dcolumn}
\usepackage{bm}
\usepackage{float}
\usepackage{color}
\usepackage{amssymb}
\usepackage{amsmath}
\usepackage[caption=false]{subfig}
\usepackage{xcolor}
\usepackage[percent]{overpic}
\usepackage{enumitem}
\usepackage{natbib}
\usepackage{overpic}
\usepackage[none]{hyphenat}

\begin{document}

\title{Cascading failures in isotropic and anisotropic spatial networks
induced by localized attacks and overloads}

\author{Ignacio A. Perez} \email{ignacioperez@mdp.edu.ar}
\affiliation{Instituto de Investigaciones F\'isicas de Mar del
  Plata (IFIMAR)-Departamento de F\'isica, FCEyN, Universidad Nacional
  de Mar del Plata-CONICET, De\'an Funes 3350, (7600) Mar del Plata,
  Argentina}
\author{Dana Vaknin Ben Porath}
\affiliation{Department of Physics, Bar-Ilan University, Ramat-Gan
  52900, Israel}
\author{Cristian E. La Rocca}
\affiliation{Instituto de Investigaciones F\'isicas de Mar del
  Plata (IFIMAR)-Departamento de F\'isica, FCEyN, Universidad Nacional
  de Mar del Plata-CONICET, De\'an Funes 3350, (7600) Mar del Plata,
  Argentina}
\author{Sergey V. Buldyrev}
\affiliation{Department of Physics, Yeshiva University,
New York 10033, USA}
\affiliation{Physics Department,
  Boston University, 590 Commonwealth Ave., Boston, Massachussets
  02215, USA}
\author{Lidia A. Braunstein}
\affiliation{Instituto de Investigaciones F\'isicas de Mar del
  Plata (IFIMAR)-Departamento de F\'isica, FCEyN, Universidad Nacional
  de Mar del Plata-CONICET, De\'an Funes 3350, (7600) Mar del Plata,
  Argentina}
\affiliation{Physics Department,
  Boston University, 590 Commonwealth Ave., Boston, Massachussets
  02215, USA}
\author{Shlomo Havlin}
\affiliation{Department of Physics, Bar-Ilan University, Ramat-Gan
  52900, Israel}
\affiliation{Physics Department,
  Boston University, 590 Commonwealth Ave., Boston, Massachussets
  02215, USA}

\begin{abstract}
  
  \noindent
  Cascading failures are catastrophic processes that can destroy the
  functionality of a system, thus, understanding their development in
  real infrastructures is of vital importance. This may lead to a
  better management of everyday complex infrastructures relevant to
  modern societies, e.g., electrical power grids, communication and
  traffic networks. In this paper we examine the Motter-Lai
  model~\cite{mott-02} of cascading failures induced by overloads in
  both isotropic and anisotropic spatial networks, generated by
  placing nodes in a square lattice and using various distributions of
  link lengths and angles. Anisotropy has not been earlier considered
  in the Motter-Lai model and is a real feature that may affect the
  cascading failures. This could reflect the existence of a preferred
  direction in which a given attribute of the system manifests, such
  as power lines that follow a city built parallel to the coast. We
  analyze the evolution of the cascading failures for systems with
  different strengths of anisotropy and show that the anisotropy
  causes a greater spread of damage along the preferential direction
  of links. We identify the {\it critical linear size}, $l_c$, for a
  square shaped localized attack, which satisfies with high
  probability that above $l_c$ the cascading disrupts the giant
  component of functional nodes, while below $l_c$ the damage does not
  spread. We find that, for networks with any characteristic link
  length, their robustness decreases with the strength of the
  anisotropy. We show that the value of $l_c$ is finite and
  independent of the system size (for large systems), both for
  isotropic and anisotropic networks. Thus, in contrast to random
  attacks, where the critical fraction of nodes that survive the
  initial attack, $p_c$, is usually below 1, here $p_c = 1$. Note that
  the analogy to $p_c = 1$ is also found for localized attacks in
  interdependent spatial networks~\cite{berez-15}. Finally, we measure
  the final distribution of functional cluster sizes and find a
  power-law behavior, with exponents similar to regular
  percolation. This indicates that, after the cascade which destroys
  the giant component, the system is at a percolation critical
  point. Additionally, we observe a crossover in the value of the
  distribution exponent, from critical percolation in a
  two-dimensional lattice for strong spatial embedding, to mean-field
  percolation for weak embedding.
  
\end{abstract}

\maketitle

\section{Introduction} 

Real-world infrastructures such as power grids, sewer networks, and
telecommunication systems can be particularly affected by a process
known as cascading failures (CF). This is a dynamic process in which
the malfunction of one or a few components of the system leads to the
failure of other components, and so on, and could cause a large
fraction of the system to collapse. Motter and Lai~\cite{mott-02}
modeled the process of CF induced by overloads. In their model, at
each time step, some relevant physical quantity (e.g., energy or
information) is exchanged between every pair of nodes and transmitted
along the shortest path~\cite{new-10,bunde-91} between them. Then, the
load $L_i$ at node $i$ is defined as the total number of shortest
paths passing through this node, and the capacity $C_i$ is the maximum
load that the node can handle. Since in man-made networks capacity is
limited by cost, Motter and Lai~\cite{mott-02} assumed that the
capacity of node $i$ is proportional to the original load $L^0_i$,
i.e., $C_i = (1 + \alpha) L^0_i$, where the constant $\alpha \ge 0$ is
the {\it tolerance} of the system to overloads. The failures of nodes
cause the redistribution of shortest paths. As a result, the load at
some nodes can increase and exceed their capacities $C_i$. These
overloaded nodes then fail and, since the shortest paths change, this
may cause further overloads and subsequent failures, in a cascade
manner~\cite{korn-18}. This kind of CF can be catastrophic, meaning that even a small
amount of initial failures can damage a substantial portion of the
system, and eventually cause its total
collapse~\cite{dobs-92,dobs-07,mott-02,carr-16}.

Usually, infrastructures are embedded in space, and thus several
models of CF in spatially-embedded networks have been proposed for
approaching this problem~\cite{zhao-16, vak-17}. Zhao et
al.~\cite{zhao-16} modeled a cascade of failures induced by overloads
in a square lattice. They triggered the cascade with a localized
attack, which usually occurs in natural catastrophes or malicious
attacks by removing nodes at the center of the network. They found
that failures spread radially from the center of the initial attack
with an approximately constant velocity, which decreases with
increasing tolerance. However, real systems are rarely found to be
perfect lattices and present, instead, a characteristic euclidean
link length $\zeta$, such as the European power grid and the inter
station local railway lines in Japan~\cite{wax-88,daq-11,2012japan}.
This feature can be modeled by assuming that links connecting
different nodes in a lattice have a link length distribution $P(r)
\sim exp(-r/\zeta)$~\cite{wax-88}, where $r$ is the Euclidean distance
between two nodes and $\zeta$ determines the typical length of links
in the spatial embedding. For $\zeta \to \infty$ all link lengths are
equally likely and spatial effects vanish, while for smaller values of
$\zeta$ the strength of the spatial embedding increases, as shorter
link lengths are favored.

Within spatial networks, a prominent characteristic of many real
infrastructures (e.g., grids, pipeline systems, and transportation
networks) is the presence of anisotropy in the orientation of the
connections between nodes. Usually, the disposition of links is not
the same in all directions, they rather follow the distribution of the
population, which can be anisotropic since, in many cases, it spreads
along geographical landscapes (e.g., rivers, sea coasts, mountain
ranges) or major transportation routes. It is expected that these
deviation from an isotropic lattice will change significantly the
characteristics of CF spread. 

In this paper we study both the critical damage size that will disrupt
a spatially-embedded network and the effects of anisotropy in the
embedding on CF induced by overloads. Without loss of generality, we
study here the case where the overload failures are triggered by a
square shaped, localized attack at the center of the network. We
introduce anisotropy as a preferential direction in which links are
more likely to form. We find that, although anisotropy hinders the
propagation of failures along the orthogonal direction, it
deteriorates the overall robustness of the system, and that the
initial critical damage size is independent of the system
size. Analyzing the distribution of functional cluster sizes, $n_s$,
at the end of the CF, we find that, as a result of the network
fragmentation, the distribution behaves like a power-law, $n_s \sim
s^{-\tau}$. Furthermore, we observe a crossover for the exponent
$\tau$, which changes from the known exponent value of critical
percolation in two-dimensional lattices, for high spatial embedding,
to the mean field value as the embedding declines.

\section{The $\zeta$-model}

We model the process of CF produced by spreading of overloads, in a
spatially-embedded and weighted network, which we call the $\zeta$-model
~\cite{dan-16}. The nodes of the network are placed in the vertices of a
square lattice of size $L \times L$. The lengths, $r$, of the links
are taken from the distribution $P(r) \sim exp(-r/\zeta)$. That is,
$\zeta$ is the characteristic length of the links. Also, we use here
rigid boundary conditions, but similar results can be obtained for
periodic boundary conditions. To model the isotropy and the
anisotropy, the directions of links (i.e., the angles $\theta$ that
links form with the horizontal axis), are taken from a uniform
distribution $U_{[0, 2\pi)}$ or a Gaussian distribution
$N(\theta_p,\sigma^2)$, where $\theta$ and $\theta + 2\pi$ correspond
to the same direction. In the anisotropic distribution,
$N(\theta_p,\sigma^2)$, the angle $\theta_p$ represents the
preferential direction for the connections, while $\sigma$ is the
corresponding standard deviation, which controls the strength of the
anisotropy. For instance, $\sigma \to 0$ represents the case of a
unique possible angle for the connections between nodes, while $\sigma
\to \infty$ corresponds to an isotropic network, where links appear
with equal probability in all directions. The weight of a link
represents, e.g., the time needed to traverse it, if the optimal path
is defined as the path with a minimal travel time. We take the
weights, which are independent of $r$, from a Gaussian distribution
$N_{\omega}(\omega^*,\sigma_{\omega}^2)$, where $\omega^*$ is the
average weight and $\sigma_{\omega}$ is the corresponding standard
deviation.

\begin{figure}[h]
  \subfloat{\begin{overpic}[width=7.0cm,height=5.0cm,angle=0]{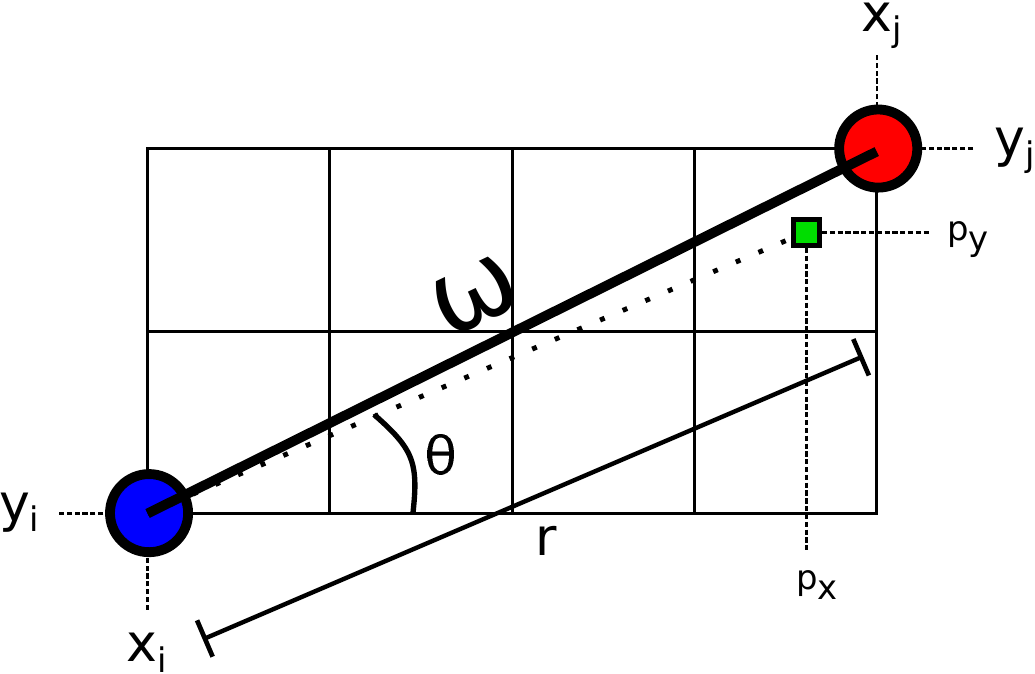}
      \put(0,60){\bf{(a)}}
  \end{overpic}} \\
  \vspace{0.5cm}
  \subfloat{\begin{overpic}[height=5.0cm]{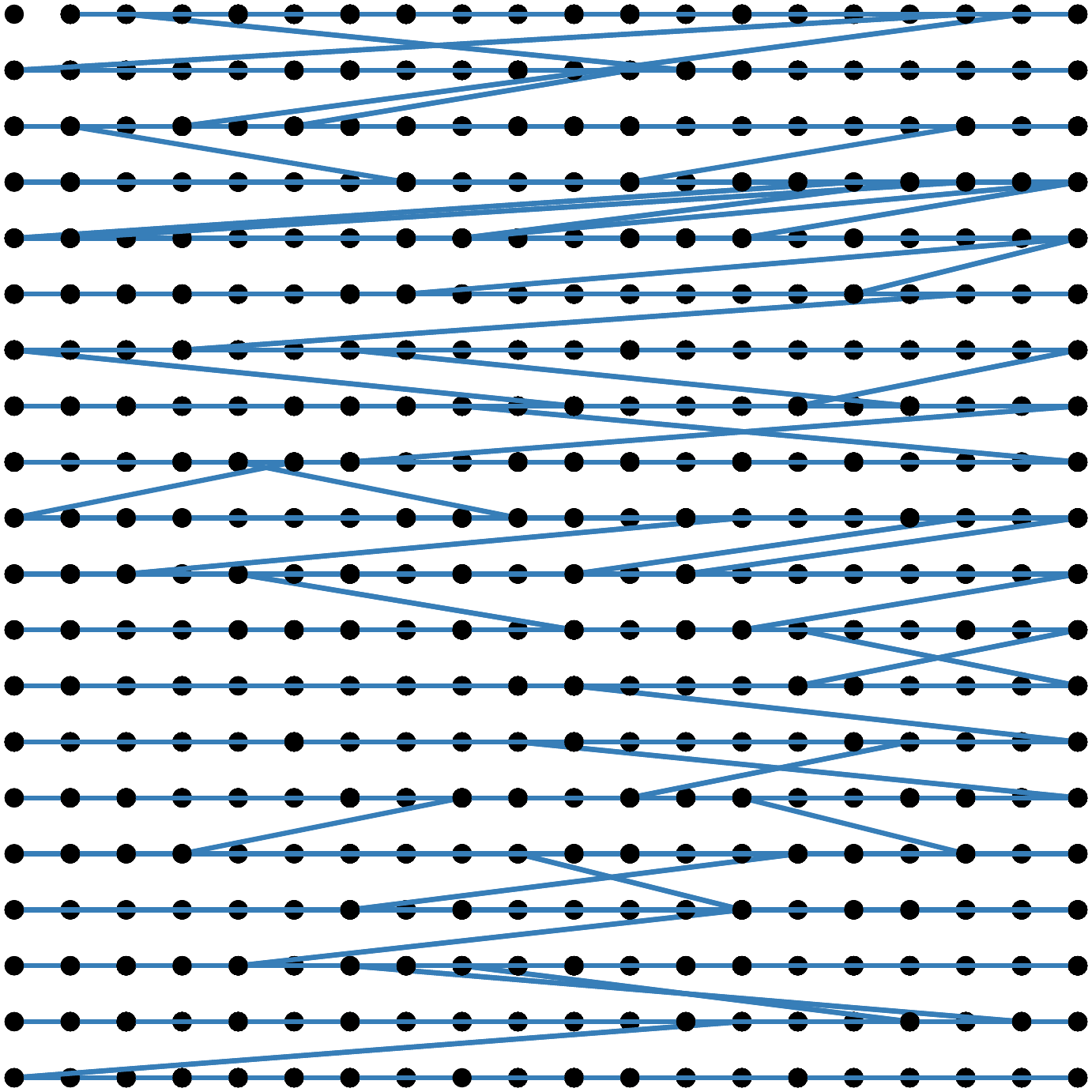}
      \put(0,105){\bf{(b)}}
  \end{overpic}}
  \hspace{0.8cm}
  \subfloat{\begin{overpic}[height=5.0cm]{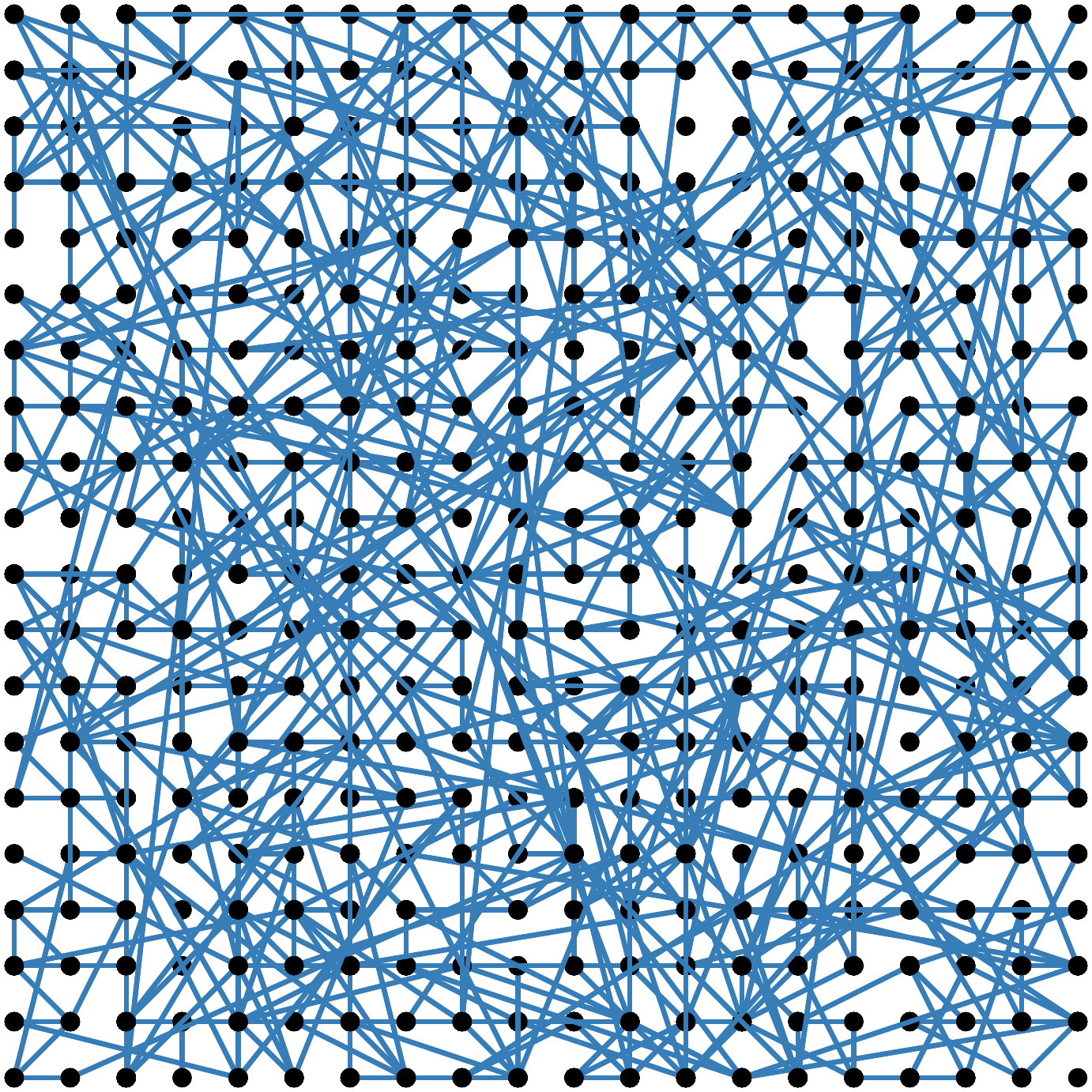}
      \put(0,105){\bf{(c)}}
  \end{overpic}}
  \caption{(a) Constructing the network's links. First, a node $i$ with
  coordinates $(x_i,y_i)$ in the lattice (blue circle) is selected
  at random. A link is created between this node and node $j$ (red
  circle), which is the closest node to the site
  $(p_x,p_y) = (x_i + r cos \theta,y_i + r sin\theta)$ marked as a green
  square. The link is represented with a straight line and is assigned a
  weight $\omega$. Nodes in the remaining sites of the lattice are not
  shown for clarity. In the bottom figures, we show two different networks
  from our model with: (b) high anisotropy ($\sigma = 0.05$) and (c)
  isotropy, both for $L = 20$ and $\zeta = 3$.}
  \label{link}
\end{figure}

To construct the $\zeta$-model network, we assign $(x,y)$ integer
coordinates $(x, y \in [1,L])$ for each of the $N = L \times L$
nodes. Then we select, at random, a node $i$ with coordinates
$(x_i,y_i)$ and draw a ray of length $r$ and angle $\theta$ above the
horizontal axis, which are randomly selected from the distributions
$P(r)$ and $N(\theta_p,\sigma^2)$, respectively. Next, we connect the
node $i$ with the node $j$ that is closest to the end point of the
ray, $p$, with real coordinates $(p_x,p_y) = (x_i + r cos \theta,y_i +
r sin\theta)$ (see Fig.~\ref{link} (a)), and assign to the link a
weight $\omega$ from the distribution
$N_{\omega}(\omega^*,\sigma_{\omega}^2)$. We repeat the process until
the total number of links in the network is $N \langle k \rangle /2$,
where $\langle k \rangle$ is the average number of links per node
(self and multiple links are not allowed). In Figs.~\ref{link} (b) and
(c), we show representations of anisotropic and isotropic networks,
respectively.

The dynamic process of CF develops as follows. Initially, at time $t =
0$, all nodes are functional, and the load at node $i$ is $L^0_i
\equiv L_i(t = 0)$, which is computed as the total number of optimal
paths~\cite{mott-02, havl-05} between all pairs of nodes that pass
through node $i$. Then, the capacity of node $i$, $C_i$, is defined as
$C_i = (1 + \alpha)L^0_i$, where $\alpha$ is the homogeneous (i.e.,
the same for all nodes) tolerance of the system to overloads. We
assume that $\alpha$, and thus $C_i$, are constants throughout the
entire process. At time $t = 1$, we generate a square shaped,
localized attack of $\Delta l = l \times l$ nodes which fail at the
center of the network. Due to this attack, the optimal paths may be
redistributed, changing the load of some nodes. Then, at time $t > 1$,
all nodes fail if their loads (which must be computed at every single
time step) are above their capacities, i.e., if $L_i(t) > C_i$. The
process continues until the failures produced at a given time step
cause no new nodes to fail at a posterior time.

It is important to note that initially, the giant component (GC) in
the network (i. e., the biggest group of nodes where each node has a
path to each other) spans the system reaching all four boundaries.

\section{Results}

Before presenting the actual results of this study, we note that in
the limit of an isotropic network ($\sigma \to \infty$) and for small
values of $\zeta$ (e.g., $\zeta = 1$), our model coincides with the
lattice model of Ref.~\cite{zhao-16}. In the Supplementary Information
section, we provide results that support this statement. For all
simulations (including those in the SI section) we use, without loss
of generality, $\langle k \rangle = 4$, $\theta_p = 0$, $\omega^* = 5$, and
$\sigma_{\omega} = 0.1$.

Now, we begin with the analysis of our model and study the evolution
of the cascade and the spatial distribution of the failures for a
fixed size of the initial attack, in networks with different degrees
of anisotropy. We denote the number of nodes that fail at time $t$ as
$F_t \equiv F(t)$, while $R_p \equiv R_p(t)$ and $R_o \equiv R_o(t)$
are the standard deviations of the coordinates of all nodes failed
during all time steps from 0 to $t$, in the preferential (horizontal)
and orthogonal directions, respectively.

\begin{figure}[h]
  \subfloat{\begin{overpic}[width=7.0cm,height=5.0cm,angle=0]{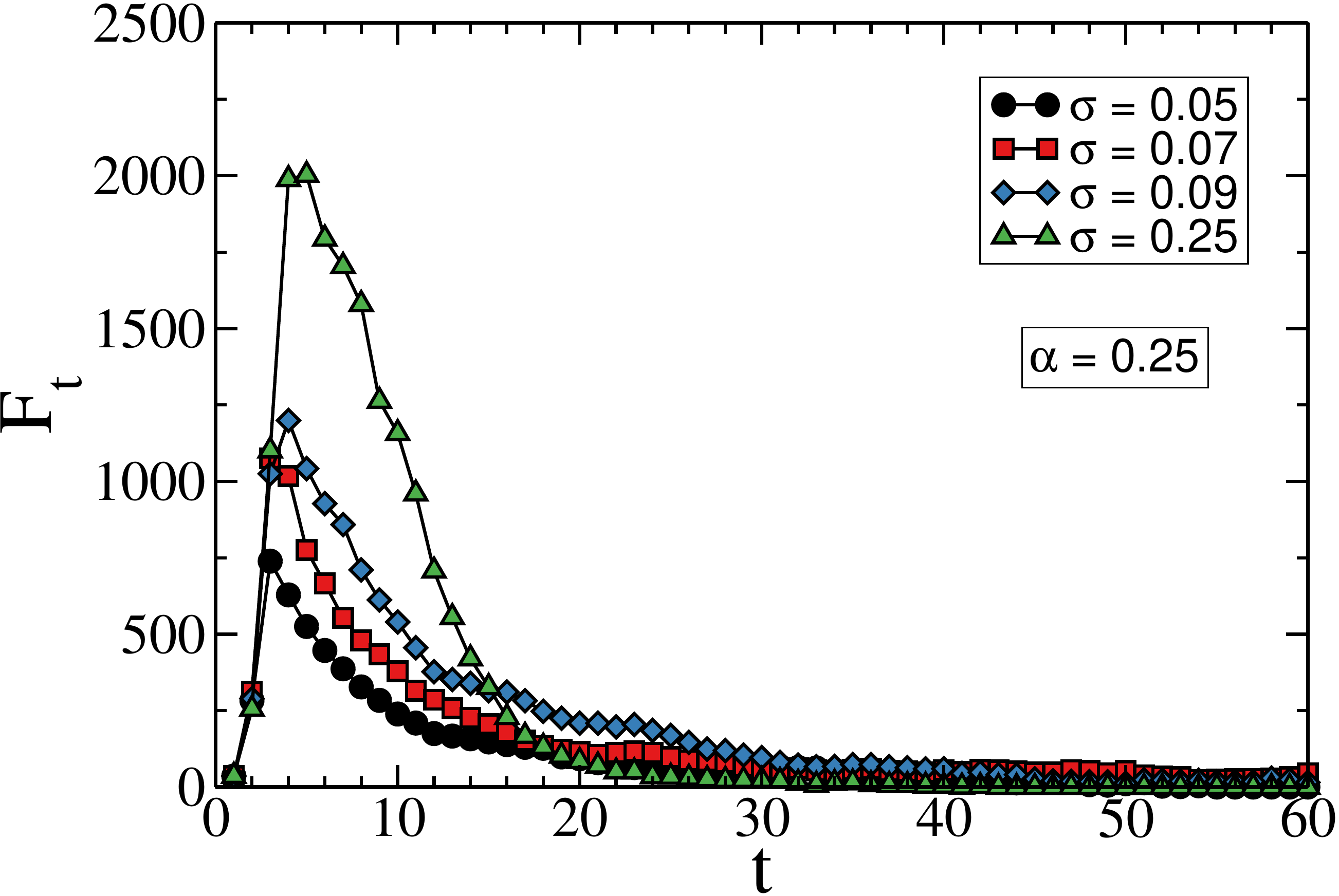}
      \put(19,62){\bf{(a)}}
  \end{overpic}}
  \subfloat{\begin{overpic}[width=7.0cm,height=5.0cm,angle=0]{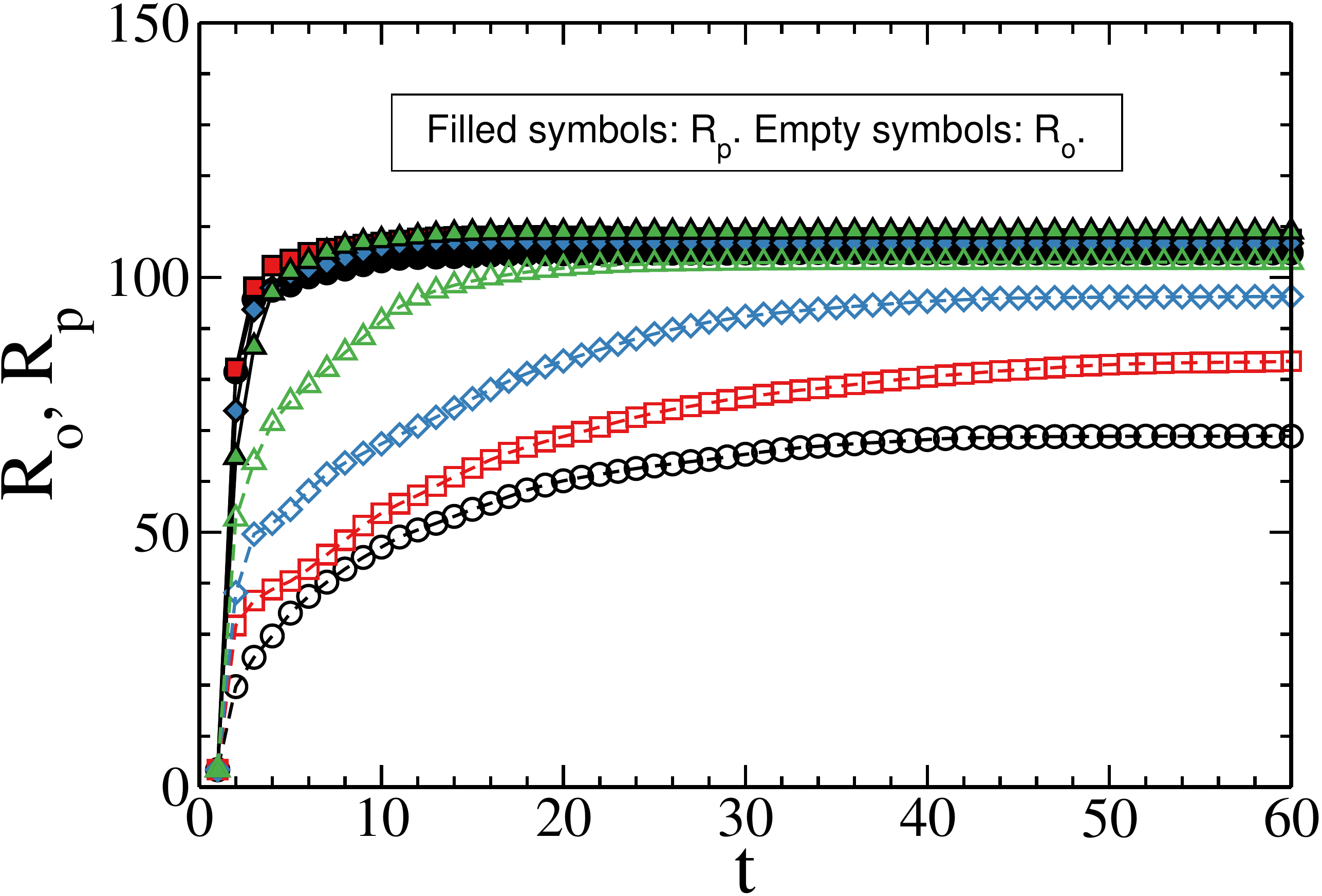}
      \put(18,62){\bf{(b)}}
  \end{overpic}} \\
  \subfloat{\begin{overpic}[width=7.0cm,height=5.0cm,angle=0]{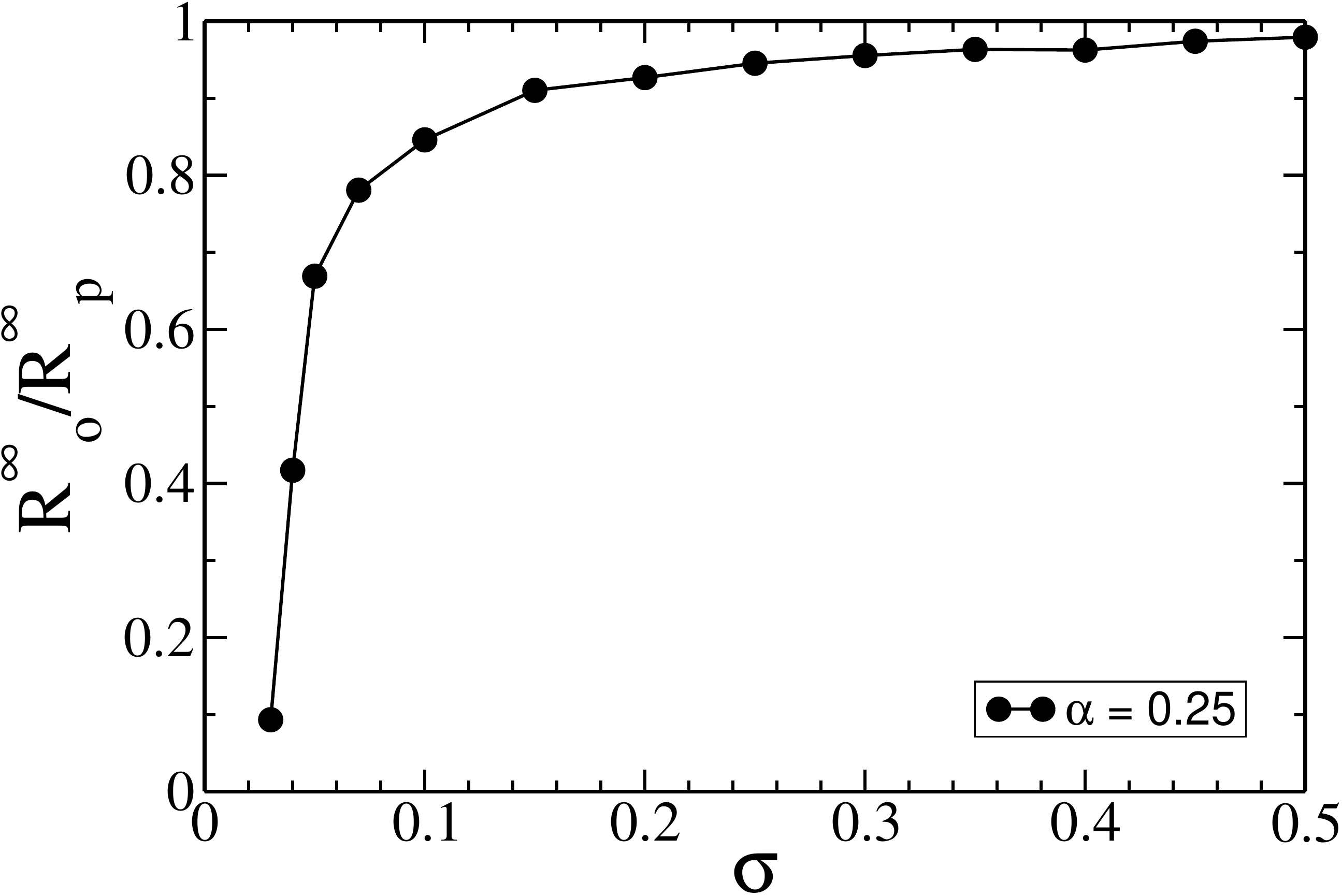}
      \put(18,62){\bf{(c)}}
  \end{overpic}}
  \hspace{1.7cm}
  \subfloat{\begin{overpic}[scale=0.32]{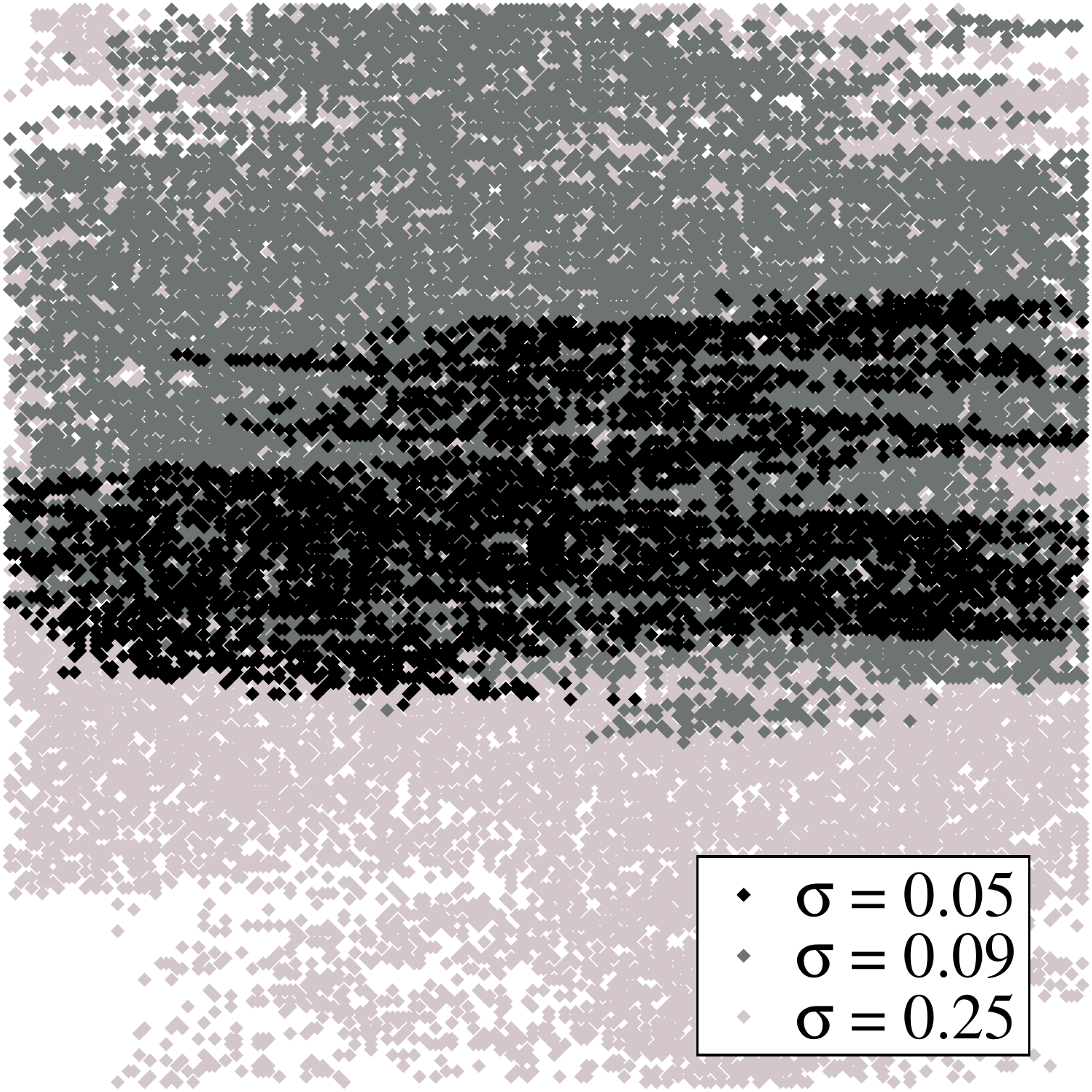}
      \put(-15,89){\bf{(d)}}
  \end{overpic}}
  \caption{(a) Number of failures, $F_t$, as a function of time $t$,
    for different values of $\sigma$. (b) Evaluating the extent of the
    failures in the preferential and orthogonal directions, $R_p$ and
    $R_o$, respectively, for the same values of the parameters shown
    in (a). (c) Ratio $R^{\infty}_o/R^{\infty}_p$ between the extent
    of the failures in each direction, at the end of the cascade. (d)
    Demonstrating individual realizations of the network showing the
    final spatial distribution of failed nodes (the exact disposition
    of failures may vary slightly between realizations). Anisotropy
    favors the propagation of failures along the preferred horizontal
    direction ($\theta_p = 0$) over the orthogonal direction. This is
    since the rare vertical links become overloaded and fail at the
    beginning of the cascade. The parameters of the simulations are
    $L = 200$, $\zeta = 3$, $\alpha = 0.25$, and $l = 6$. Results in
    (a), (b), and (c), have been averaged over $N_{rea} = 35$
    realizations. Note that all networks have initially (before the
    localized attack) a GC that spans the entire system, all four edges.}
  \label{Fr-y-Radius-z3} 
\end{figure}

In Figs.~\ref{Fr-y-Radius-z3} (a) and (b), we show $F_t$, $R_p$, and
$R_o$ for networks with $L = 200$, $\zeta = 3$, $\alpha = 0.25$, $l =
6$, and different values of $\sigma$.
We observe that the initial attack produces a CF in the network, where
the maximum amount of failing nodes at a certain time $t$ - the peak
value of $F_t$ in Fig.~\ref{Fr-y-Radius-z3} (a) - increases with
increasing $\sigma$, i.e., as the strength of the anisotropy
decreases. In this way, fewer nodes will remain functional due to
overloads, at the end of the cascading, in the isotropic case, as
compared to the anisotropic case. Also, anisotropy affects the spatial
distribution of failures over time, limiting the propagation in the
orthogonal direction ($\theta = \pi/2$) compared to that in the
preferred direction ($\theta_p = 0$), i.e., $R_o < R_p$ (see
Fig.~\ref{Fr-y-Radius-z3} (b)). However, as $\sigma$ increases and the
network loses its characteristic anisotropy, the gap between the
extents along the two directions becomes smaller.

\begin{figure}[h]
  \begin{center}
    \includegraphics[width=7.0cm,height=5.0cm,angle=0]{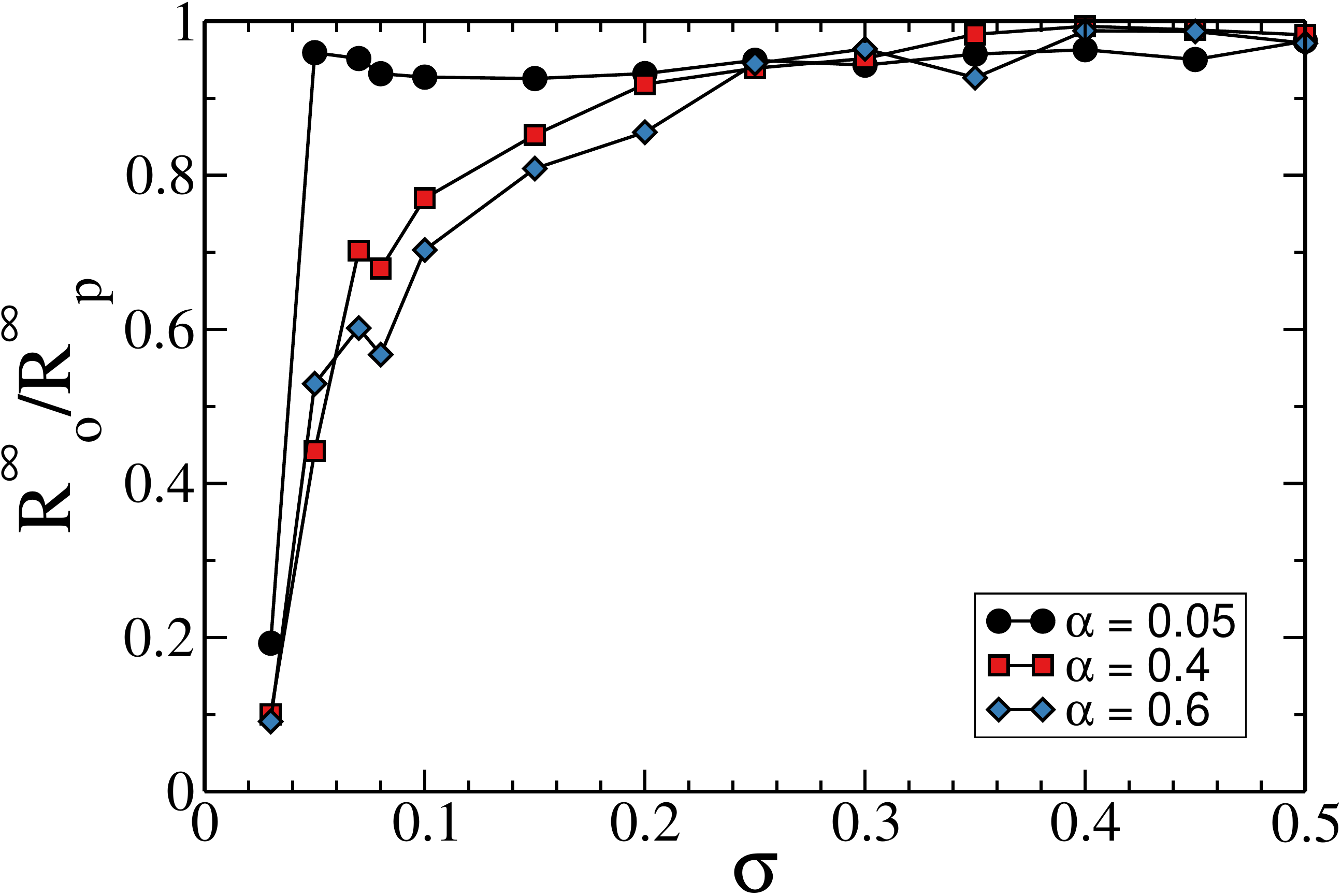}
  \end{center}
  \caption{{\bf Ratio $R^{\infty}_o/R^{\infty}_p$.} We show the ratio
    between the extents of the failures in the orthogonal ($\theta =
    \pi/2$) and preferential ($\theta_p = 0$) directions, at the end
    of the cascade, for the same networks as in
    Fig.~\ref{Fr-y-Radius-z3} (c), but for different values of the
    tolerance $\alpha$. We observe that the decay of
    $R^{\infty}_o/R^{\infty}_p$, in the limit of completely
    anisotropic networks ($\sigma \to 0$), is more abrupt as $\alpha$
    decreases since all nodes are more susceptible to
    fail.}\label{R_ratio}
\end{figure}

Additionally, in Figs.~\ref{Fr-y-Radius-z3} (c) and (d), we plot the
ratio $R^{\infty}_o/R^{\infty}_p$ ($R^{\infty} \equiv R(t \to \infty)$) and
the final distribution of failures in the network, respectively, which
help in clarifying the effects of anisotropy at the end of the cascade
process. In Fig.~\ref{Fr-y-Radius-z3} (c), we observe a convergence
towards a final isotropic state ($R^{\infty}_o/R^{\infty}_p \to 1$) for
large values of $\sigma$. For high anisotropy (small $\sigma$), the
observed ratio approaches zero and the damage does not spread
vertically. This is since the shortage of vertical links makes them to
break at early stages due to overloads (see Fig.~\ref{R_ratio} of the
SI section, where we show the same behavior for other $\alpha$
values). Finally, in Fig.~\ref{Fr-y-Radius-z3} (d) we depict examples
of the spread of failures for different degrees of anisotropy and
particular realizations of the cascade process. This gives us a visual
image of the total damage caused by the initial attack, and how it is
reduced in the orthogonal direction as the anisotropy increases (i.e.,
as $\sigma$ decreases).

In light of these results, one might be tempted to conclude that
systems with a structural anisotropy are more robust against localized
attacks, compared to isotropic systems. However, measures such as the
total amount of failures or their spatial extension might not be the
most relevant for revealing the effects that anisotropy produces in the
systems that we are exploring. In particular, note that in
Fig.~\ref{Fr-y-Radius-z3} (d), while the damage is the smallest for
$\sigma = 0.05$, the system is no longer functional since the giant
component is vertically broken.

\begin{figure}[h]
  \subfloat{\begin{overpic}[width=7.0cm,height=5.0cm,angle=0]{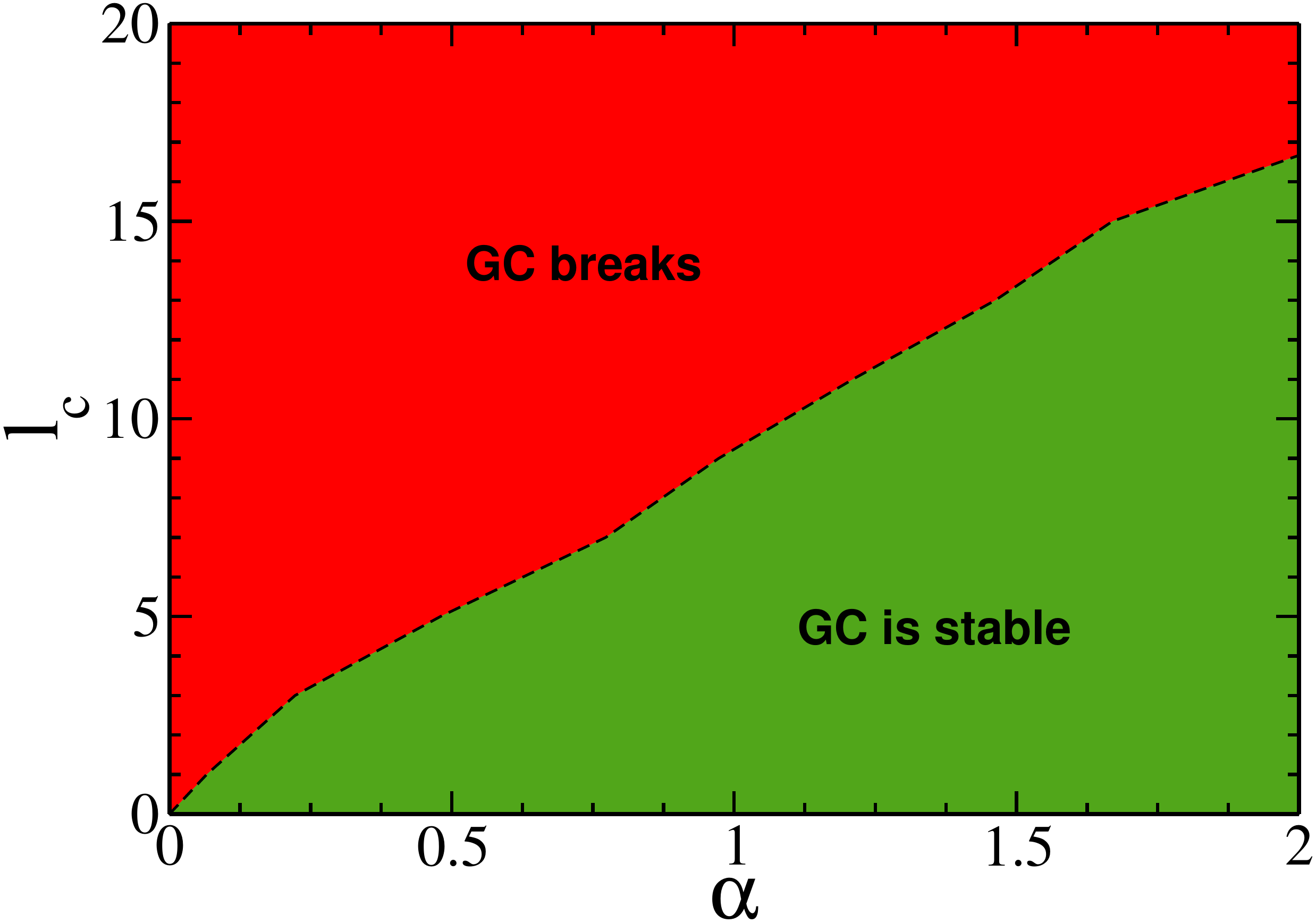}
      \put(87,13){\bf{(a)}}      
  \end{overpic}}
  \subfloat{\begin{overpic}[width=7.0cm,height=5.0cm,angle=0]{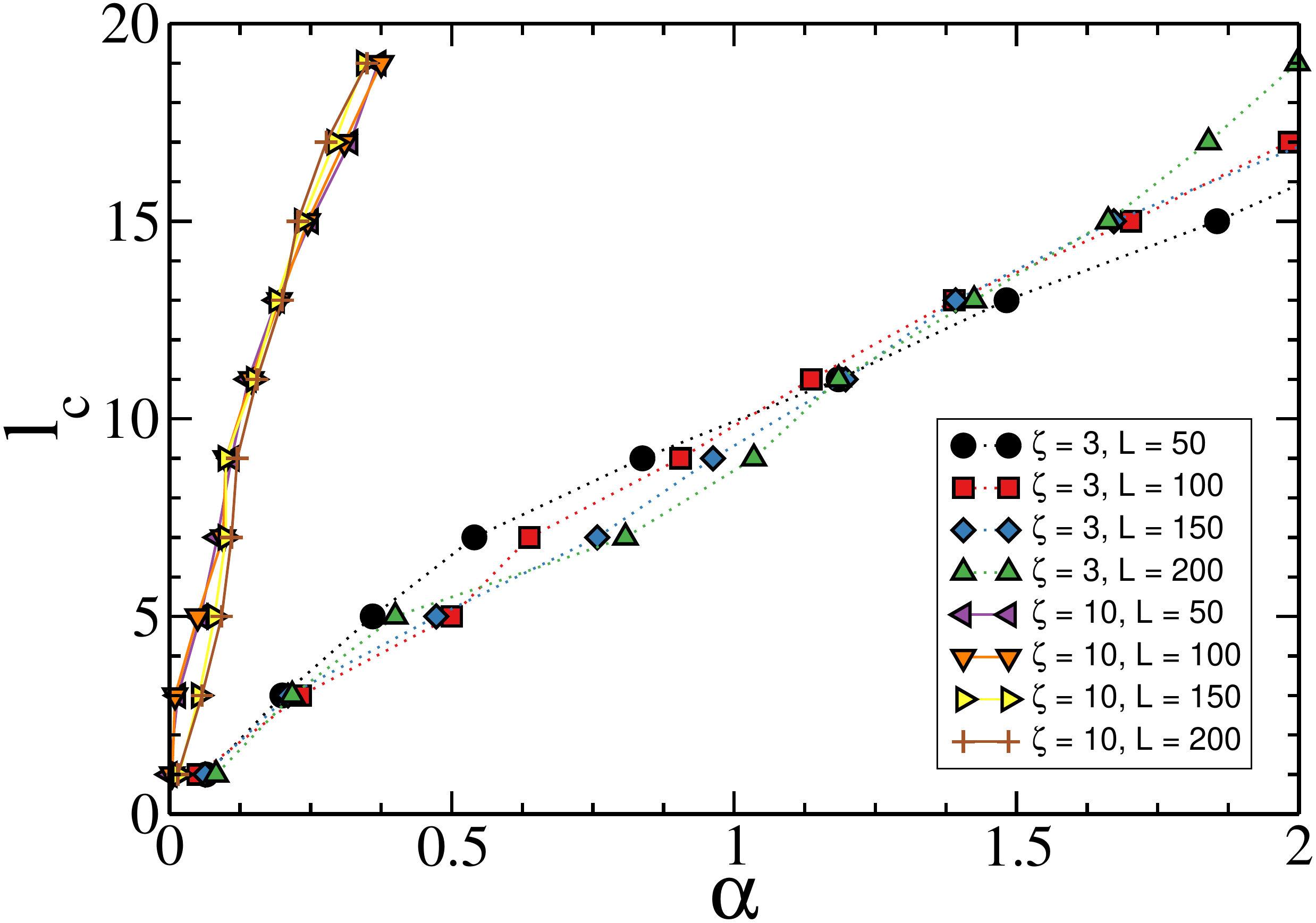}
      \put(36,60){\bf{(b)} Isotropic}
  \end{overpic}} \\
  \subfloat{\begin{overpic}[width=7.0cm,height=5.0cm,angle=0]{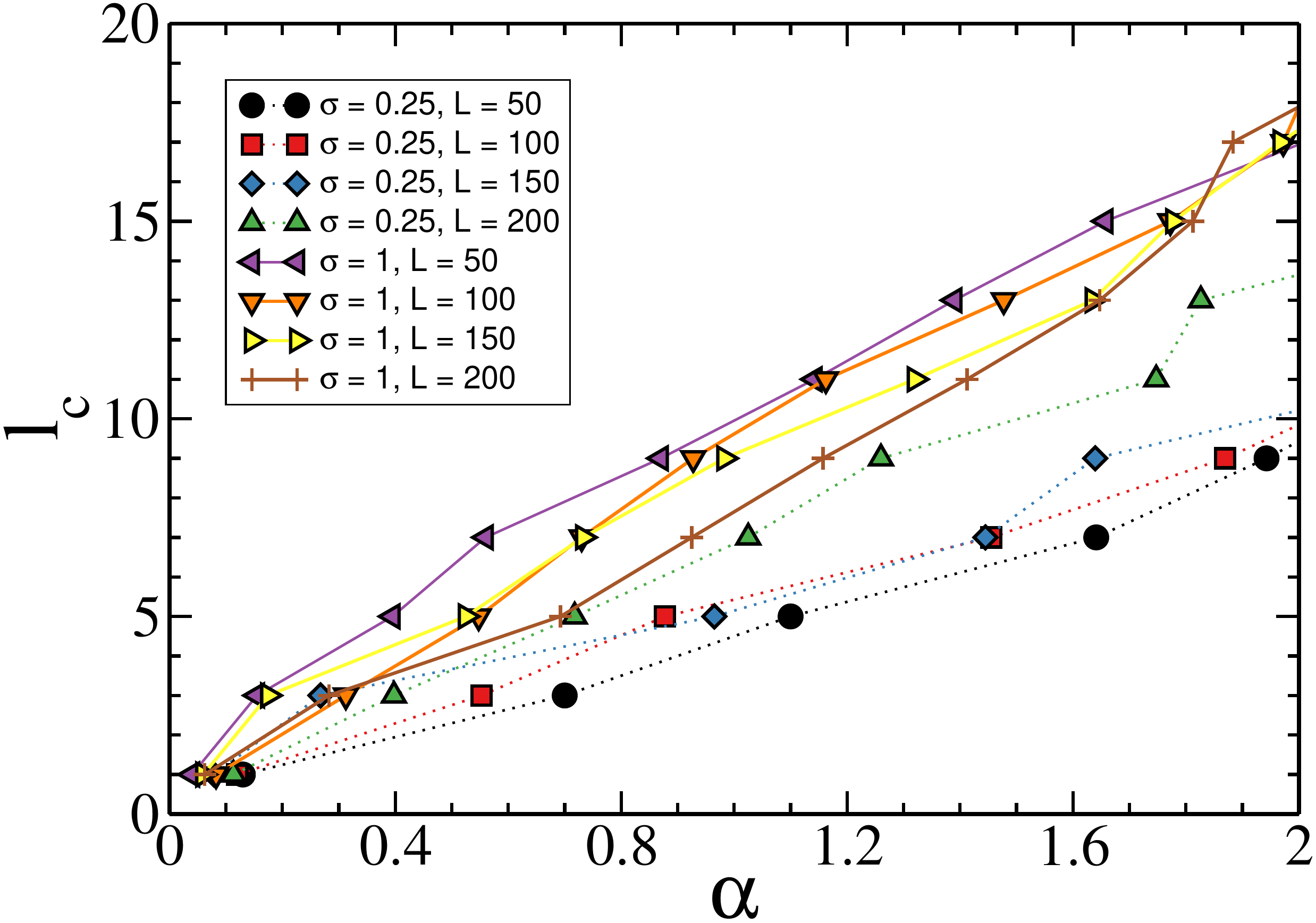}
      \put(72,13){\bf{(c)} $\zeta = 3$}
    \end{overpic}}
  \subfloat{\begin{overpic}[width=7.0cm,height=5.0cm,angle=0]{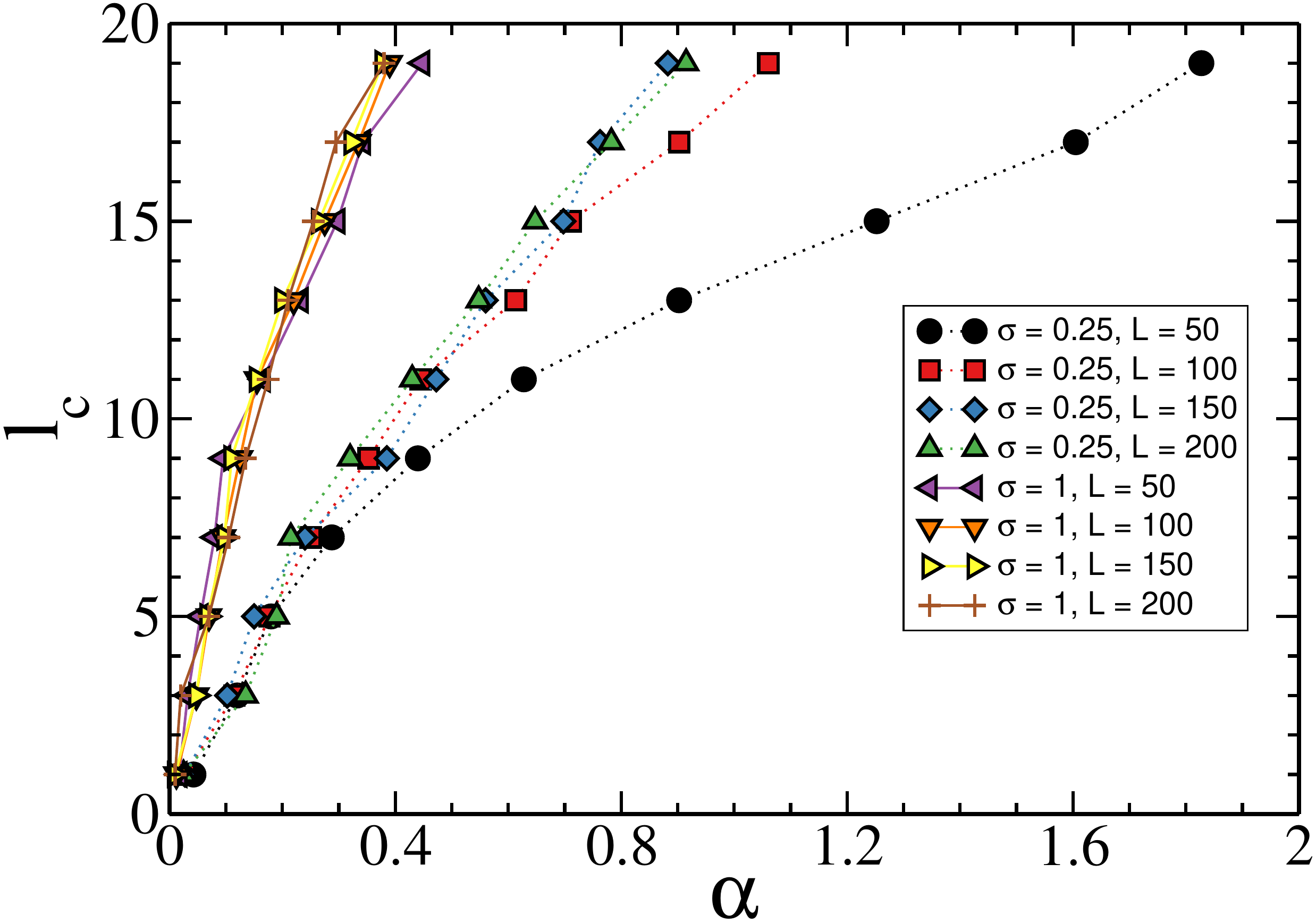}
      \put(68,13){\bf{(d)} $\zeta = 10$}
    \end{overpic}}
    \caption{Robustness of the system against localized attacks. (a)
      Scheme of a critical curve for the linear attack size $l_c$ as
      function of the tolerance $\alpha$. Below the curve (green
      region), the GC holds stable and the damage is only
      localized. Above the curve (red region), the functional GC
      breaks down, i.e., it can not reach, at least, one of the
      network edges (isotropic networks) or it can not extend
      throughout all the vertical length of the system (anisotropic
      networks). (b) Critical linear size of attack $l_c(\alpha)$ for
      isotropic networks with linear sizes $L = 50, 100, 150, 200$, and
      characteristic link lengths $\zeta = 3, 10$. Figures (c) and (d)
      show $l_c(\alpha)$ for anisotropic networks with $\zeta = 3$ and
      $\zeta = 10$, respectively, and for $\sigma = 0.25, 1$. The
      existence of a preferred direction for the links ($\theta_p =
      0$), which causes the shortage of vertical connections, impairs
      the robustness of the system, since $l_c$ decreases as $\sigma$
      decreases. The spatial embedding also debilitates the system and
      makes it easier to collapse when links are shorter. Note that
      the critical linear size $l_c$ does not depend on the system
      length $L$, both for isotropic and anisotropic networks, with
      the exception of finite size effects for $L = 50$. These results
      were obtained by computing, by means of a binary search, the
      curve $\alpha_c(l)$ from an average over $N_{rea} = 20$
      realizations, and then inverting this curve (see
      Fig.~\ref{lc-vs-alpha-w-disp} in the SI section for the
      dispersion of these values).}
    \label{lc-vs-alpha}
\end{figure}

\begin{figure}[h]
  \subfloat{\begin{overpic}[width=7.0cm,height=5.0cm,angle=0]{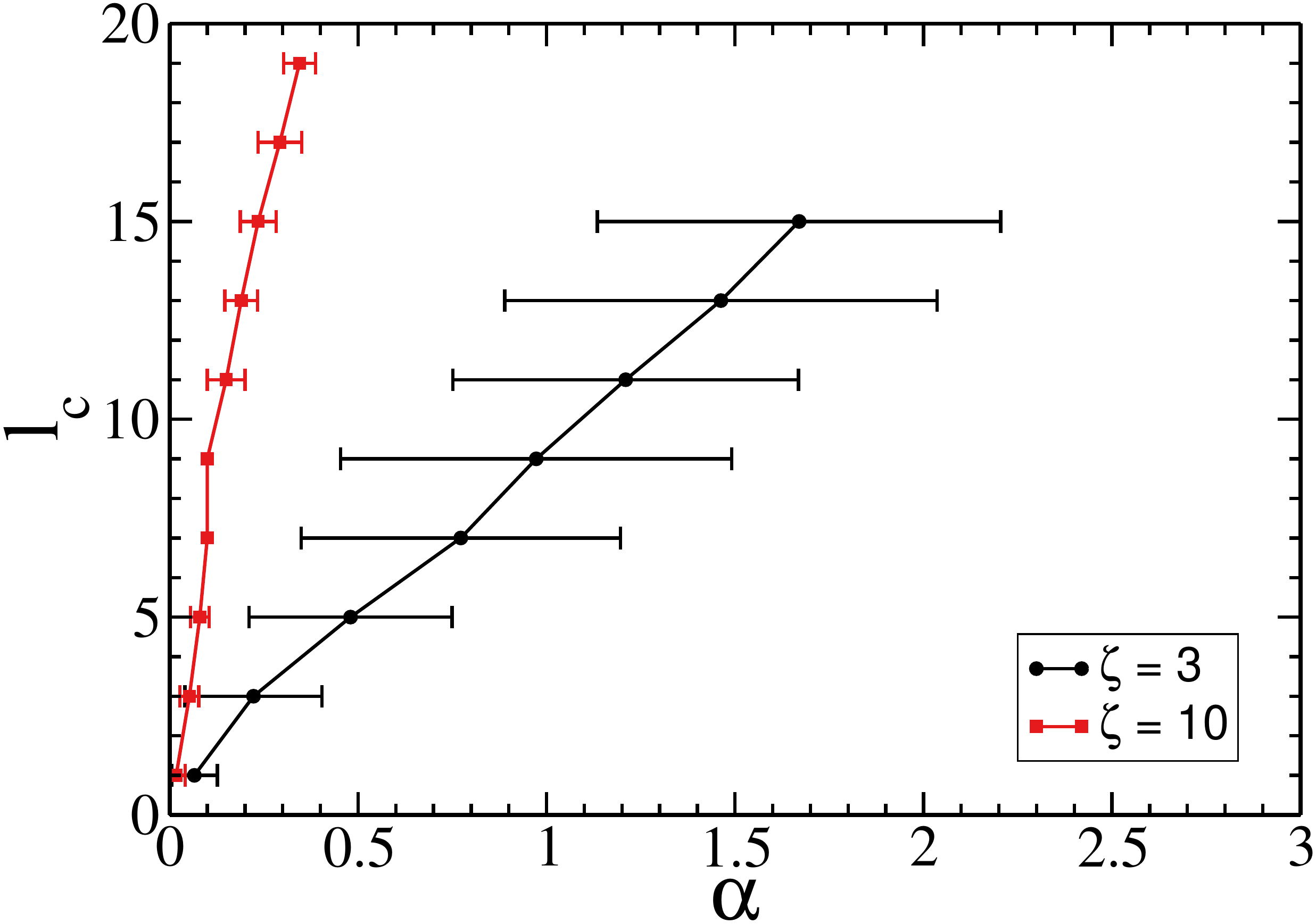}
      \put(36,60){\bf{(a)} Isotropic}
  \end{overpic}} \\
  \subfloat{\begin{overpic}[width=7.0cm,height=5.0cm,angle=0]{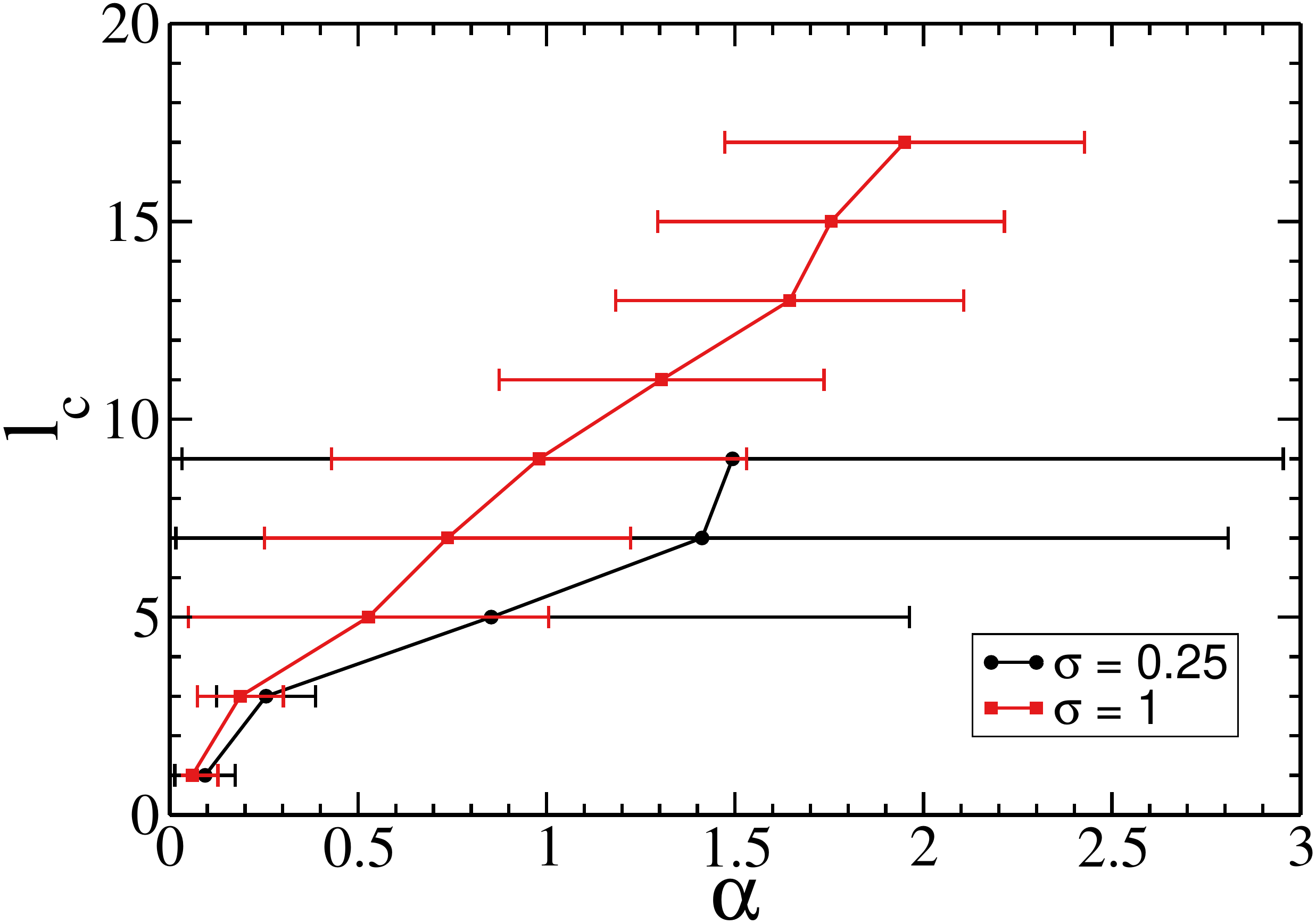}
      \put(17,60){\bf{(b)} $\zeta = 3$}
    \end{overpic}}
  \subfloat{\begin{overpic}[width=7.0cm,height=5.0cm,angle=0]{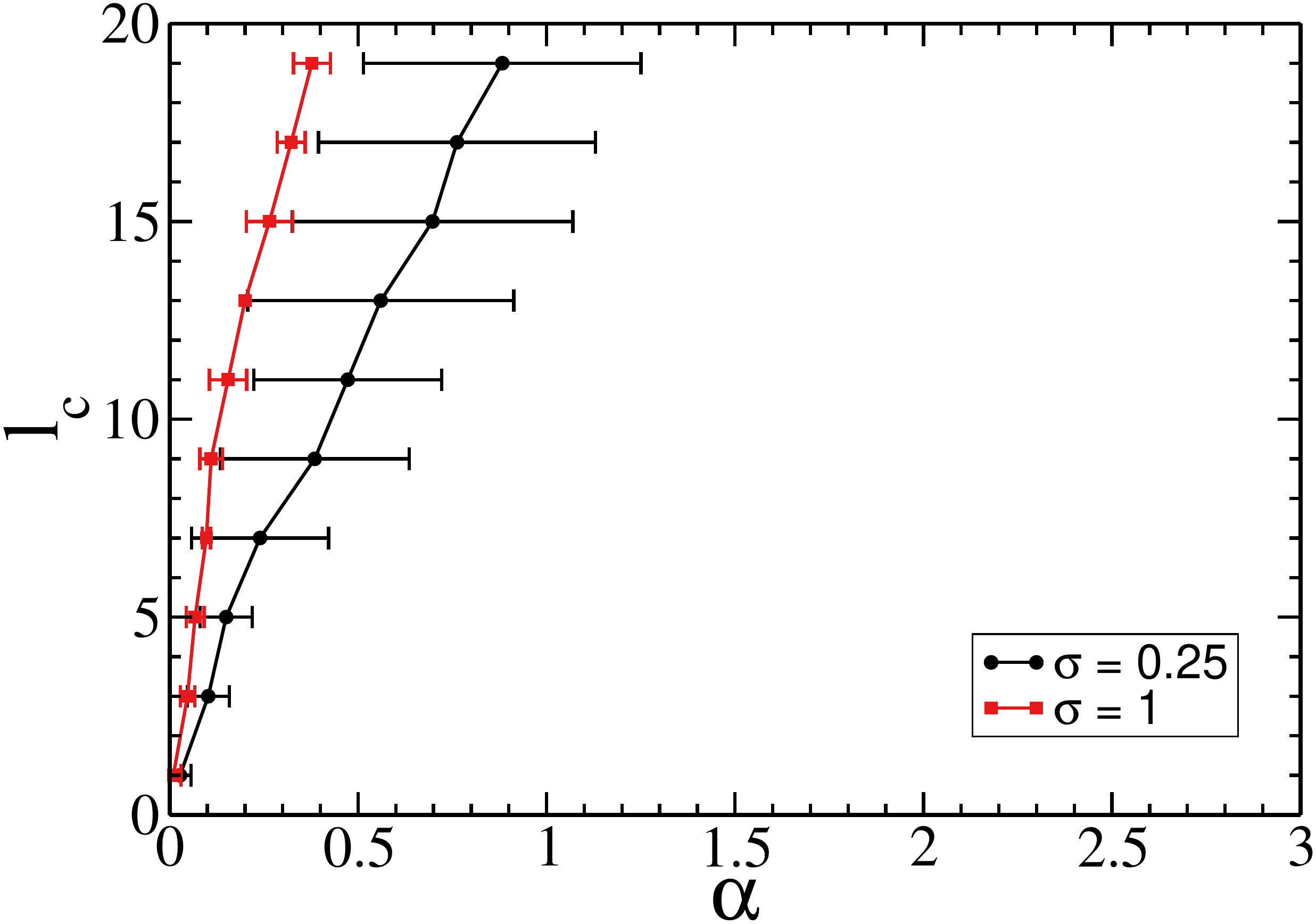}
      \put(69,60){\bf{(c)} $\zeta = 10$}
    \end{overpic}}
  \caption{{\bf Standard deviation for $l_c(\alpha)$}. We show the
    dispersion along the horizontal axis since we computed, first,
    $\alpha_c(l)$ and then inverted the curve. (a) Isotropic networks
    with $\zeta = 3, 10$. (b)-(c) Anisotropic networks with $\zeta =
    3$ (left) and $\zeta = 10$ (right), for $\sigma = 0.25, 1$. All
    the results correspond to different realizations of networks with
    linear size $L = 150$.}
    \label{lc-vs-alpha-w-disp}
\end{figure}

Next we ask what is the critical linear size of attack $l_c$, for a
given tolerance $\alpha$, such that above this value the GC of
functional nodes, at the end of the cascade, breaks down (red region
in Fig.~\ref{lc-vs-alpha} (a)), and below it the damage is only
localized (green region in Fig.~\ref{lc-vs-alpha} (a)). We define
the critical condition for computing $l_c$ as follows. For isotropic
systems, $l_c$ defines the limit where the GC, above this value, does
not reach, at least, one of the four edges of the network. For
anisotropic systems, attacks with $l > l_c$ will break the GC in the
vertical direction, i.e., the GC will not reach, at least, the upper
edge or the bottom edge of the network. We make this distinction
because, in anisotropic networks, links are more abundant along the
preferred direction ($\theta_p = 0$ in this case) than along the
perpendicular direction. Therefore, it is more likely for the GC to
break apart due to failures that propagate vertically throughout the
system. It is worth mentioning that, in order to get the values of
$l_c(\alpha)$, we generate 20 realizations of networks and, for each
attack size $l$, we use a binary search to detect the critical
tolerance, $\alpha_c$. Then, we average the results of $\alpha_c$ for
the different networks (we show dispersion values in
Fig.~\ref{lc-vs-alpha-w-disp}, in the SI section), obtaining the curve
$\alpha_c(l)$, which we adjust to present as
$l_c(\alpha)$. Accordingly, the probability of complete destruction of
the GC sharply increases near $l_c$.

First we present, in Fig.~\ref{lc-vs-alpha} (b), the critical linear
size of attack $l_c (\alpha)$ for the isotropic case and different
values of the system size $L$ and the characteristic link length
$\zeta$. For attack sizes that are below these curves, i.e., $l <
l_c(\alpha)$ (recall Fig.~\ref{lc-vs-alpha} (a)), the GC spans the
entire system, both along the preferred (horizontal) and the
orthogonal direction, while above $l_c(\alpha)$ it does not, i.e., the
GC breaks and can not reach, at least, one edge of the network. We can
see that the critical size $l_c$ increases with
the tolerance $\alpha$, since nodes with larger capacities are less
likely to become overloaded and fail, thus making the network more
robust against the attacks. It is interesting to note that the
critical size $l_c$ does not depend on the system linear size, $L$, in
the limit $L \to \infty$. Thus a zero fraction (microscopic) of
localized failures yields a macroscopic transition. This is analogous
to a similar behavior found in percolation of interdependent spatial
networks~\cite{berez-15, vaknin-20}, and is in marked contrast with
random failures for which the removal of a finite fraction of the
system is needed to cause full collapse. We also observe that the
spatial embedding triggers and enhance the propagation of the failures
in the system. Note that the critical initial damage size decreases
with decreasing the typical link length $\zeta$, thus smaller attack
sizes can trigger the more spatial systems to collapse (see
Fig.~\ref{lc-vs-alpha}).

\begin{figure}[h]
  \subfloat{\begin{overpic}[width=7.0cm,height=5.0cm,angle=0]{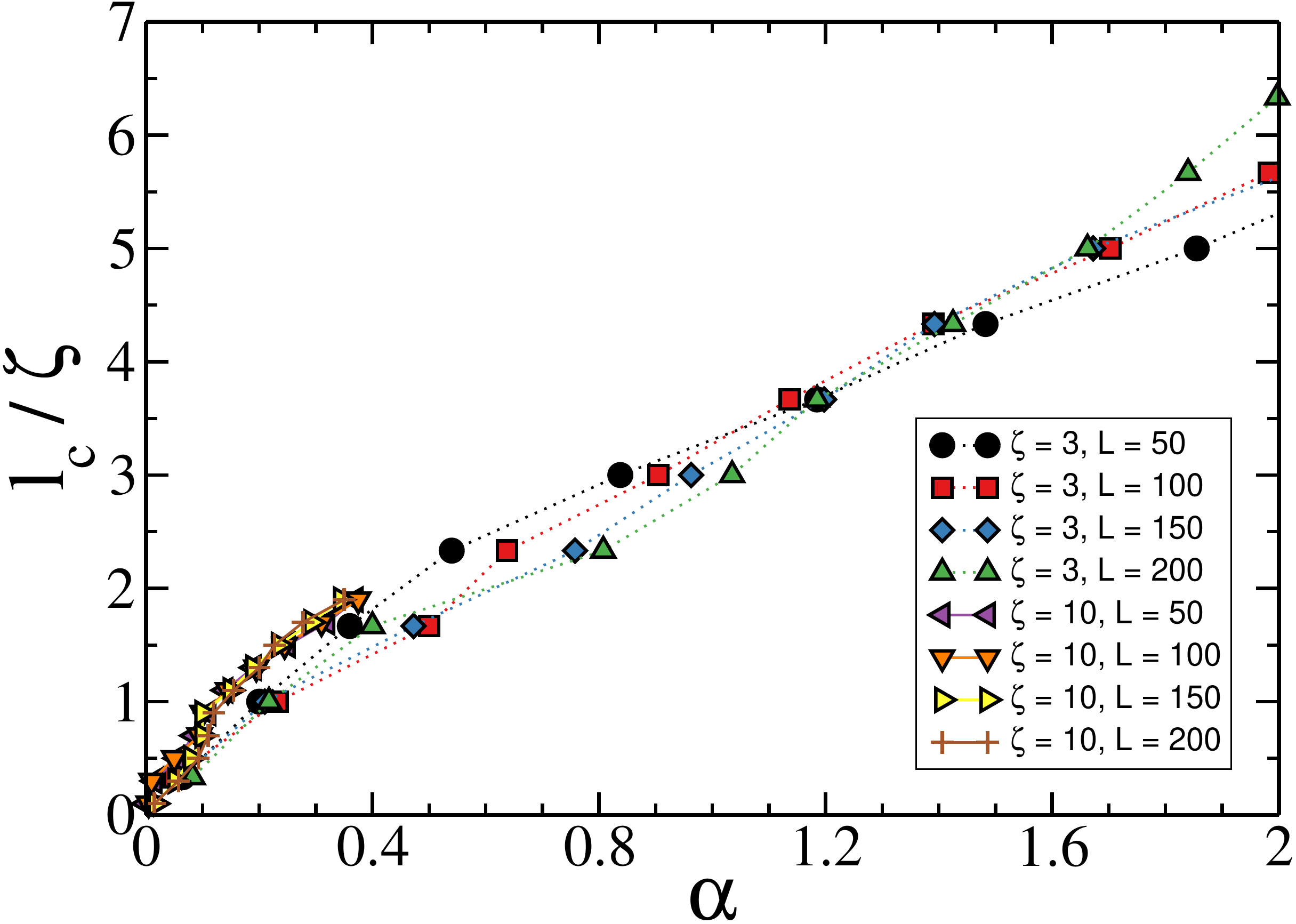}
      \put(36,60){\bf{(a)} Isotropic}
  \end{overpic}} \\
  \subfloat{\begin{overpic}[width=7.0cm,height=5.0cm,angle=0]{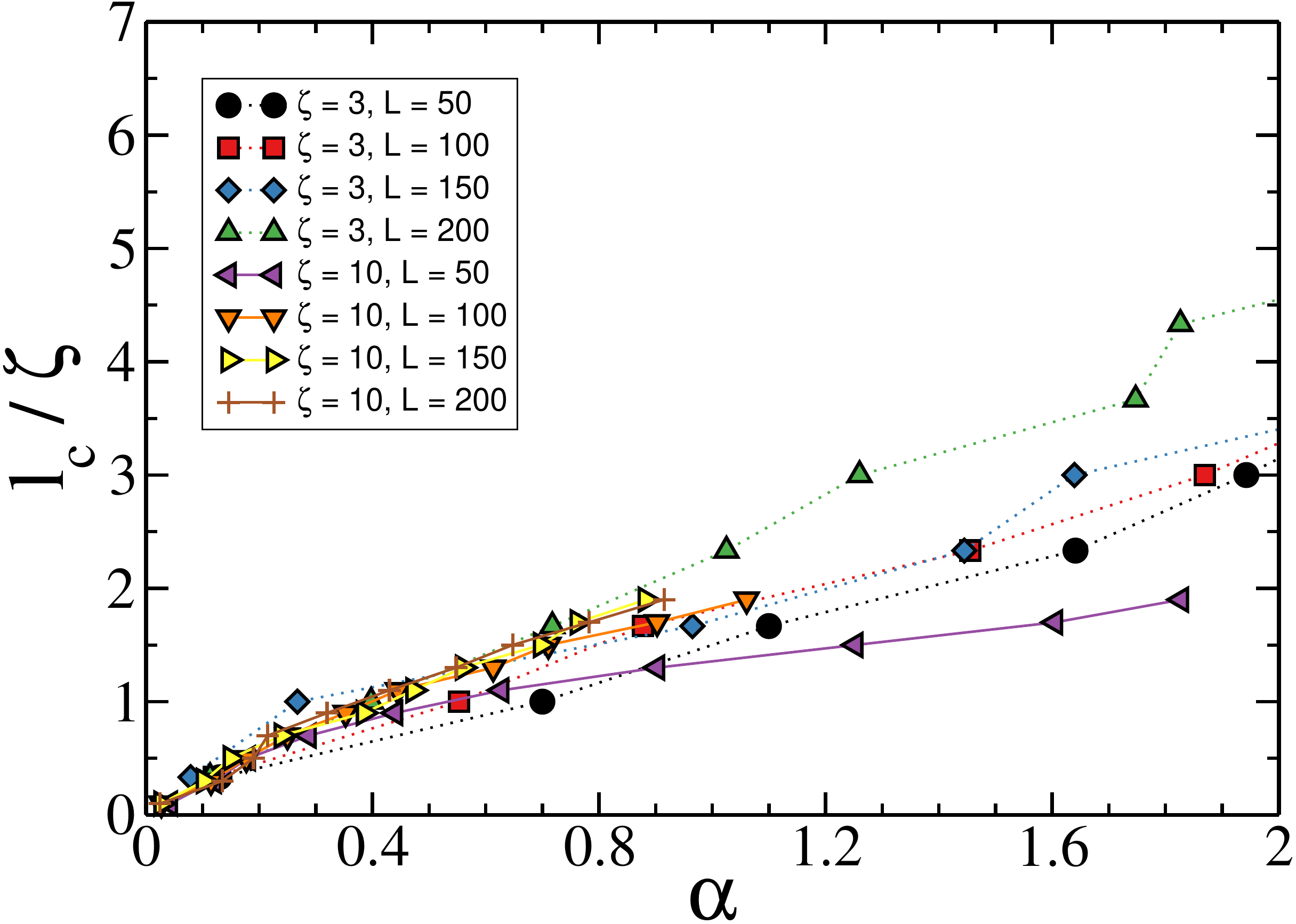}
      \put(60,13){\bf{(b)} $\sigma = 0.25$}
    \end{overpic}}
  \subfloat{\begin{overpic}[width=7.0cm,height=5.0cm,angle=0]{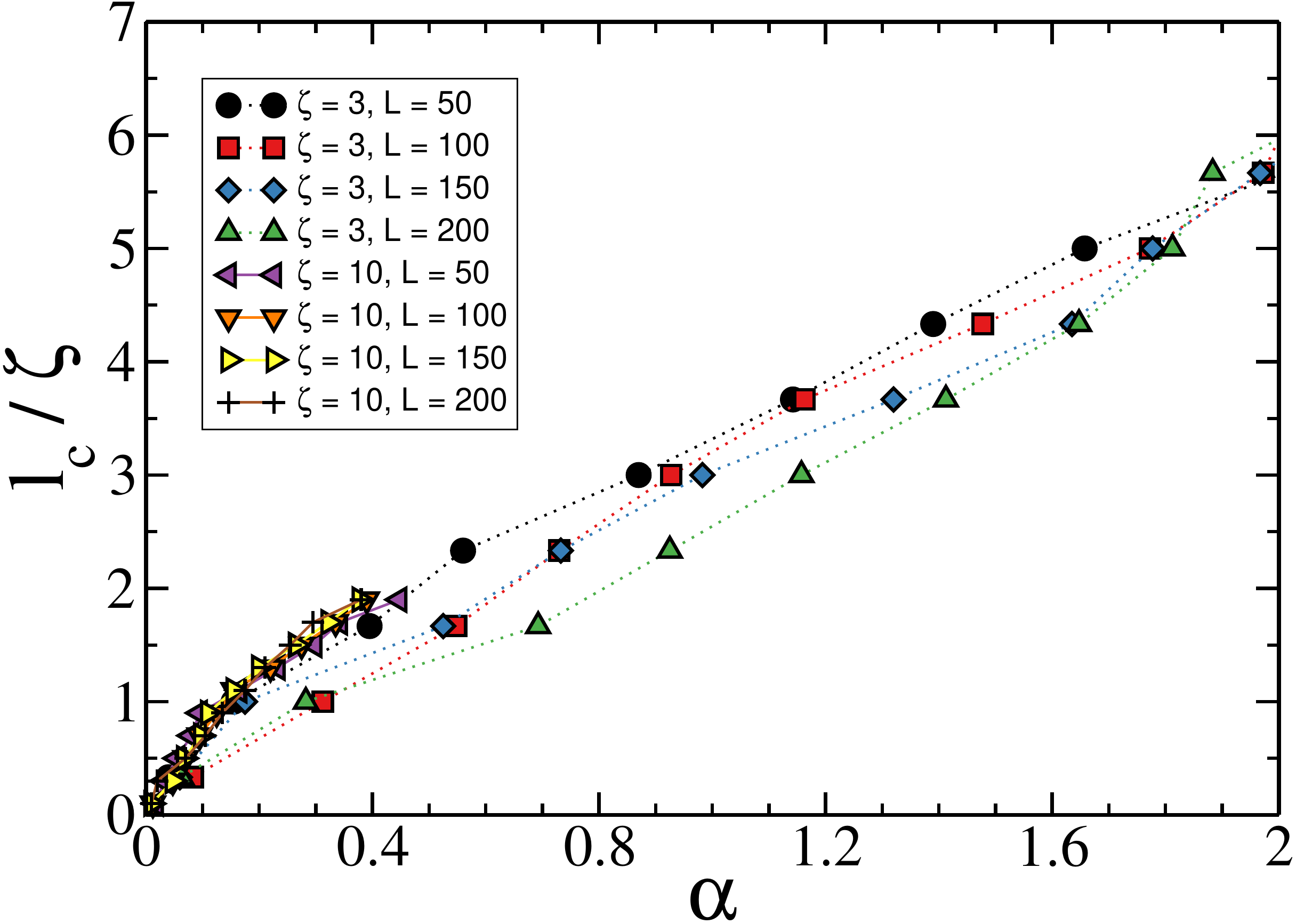}
      \put(68,13){\bf{(c)} $\sigma = 1$}
    \end{overpic}}
    \caption{{\bf The critical behavior for the scaled $l_c$.} Here we take
    results presented in Fig.~\ref{lc-vs-alpha} and divide the critical attack
    length $l_c$ by the characteristic link-length $\zeta$. The curves for the
    different values of $\zeta$ appear to approximately collapse, which suggests
    that the adimensional quotient $l_c/\zeta$ rules the behavior of the system at the
    criticality. The deviation from a single curve might be due to not enough
    statistics and due to small systems like $L = 50$. Therefore, more extensive
    computations should be carried away to assess this properly, e.g., attacks with
    $l > 20$ should be explored in networks with $\zeta = 10$ in order to get results
    for bigger $\alpha$ values, as well as increase the amount of realizations and the
    system size $L$.}
    \label{lc-over-zeta}
\end{figure}

\begin{figure}[h]
  \begin{center}
    \includegraphics[width=7.0cm,height=5.0cm,angle=0]{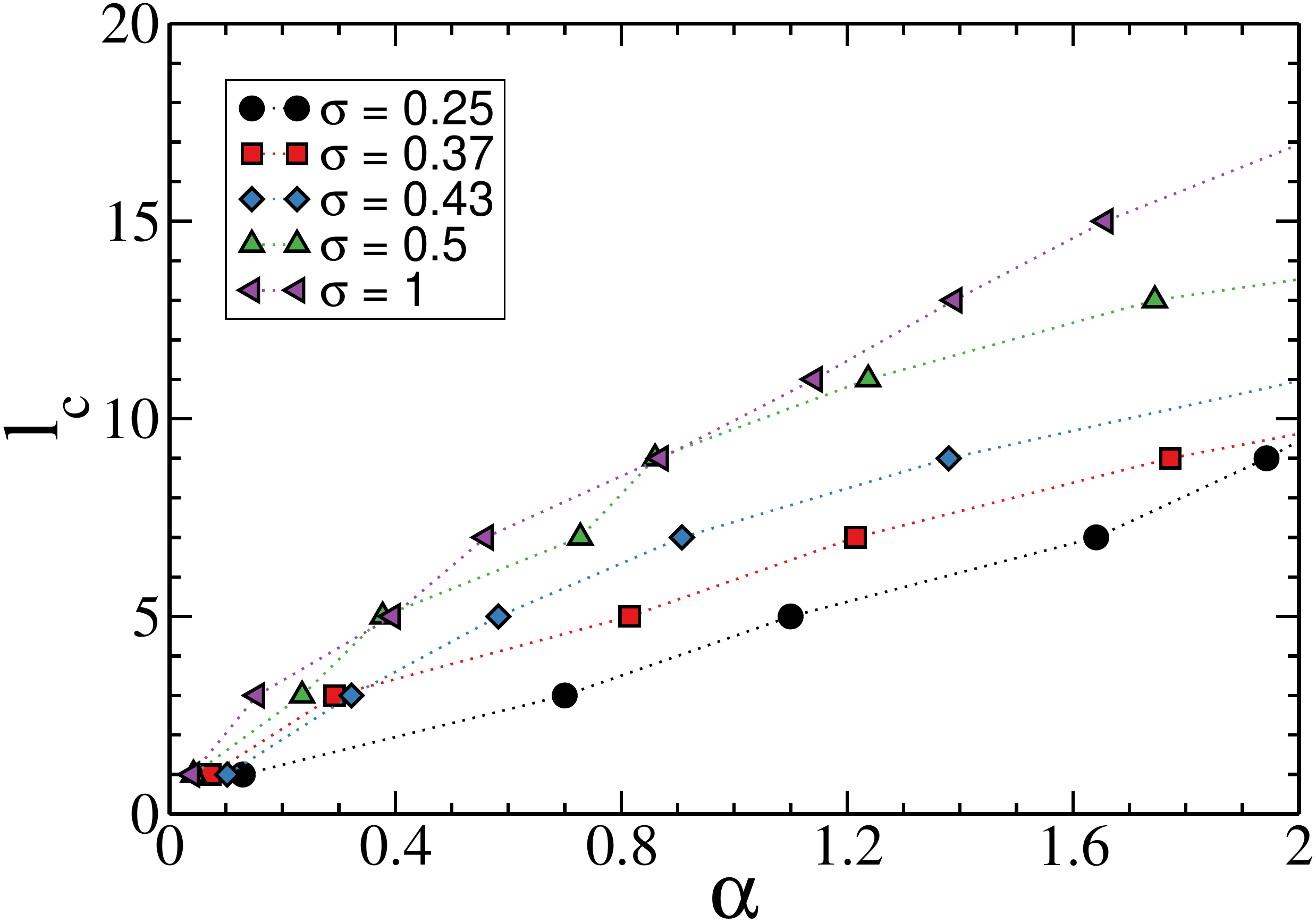}
  \end{center}
  \caption{{\bf Variation of $l_c(\alpha)$ with $\sigma$}. We present
  here the critical attack length $l_c(\alpha)$ for different values of
  the anisotropy parameter $\sigma$, in networks with $L = 50$ and
  $\zeta = 3$ (see Fig.~\ref{lc-vs-alpha} (c)). These curves suggest that
  the changes introduced by the anisotropy are gradual, taking the critical
  values to zero as $\sigma$ decreases and reaches a minimum value, for which 
  there is no GC in the system due to the lack of vertical connections.}
  \label{gradual-sigma}
\end{figure}

Next, we analyze anisotropic systems, for which results are shown in
Figs.~\ref{lc-vs-alpha} (c) and (d), and correspond to $\zeta = 3$ and
$\zeta = 10$, respectively. We show results for two different values
of the anisotropy parameter, $\sigma = 0.25, 1$, in order to depict
networks with rather different degrees of anisotropy. Note that, in
this case, attack sizes above the critical value $l_c$ produce an
horizontal damage that disrupts the flow in the vertical direction of
the network. That is, the GC does not extend between the top and
bottom edges of the system. Regardless of the spatial embedding, we
find that the critical attack size $l_c$ decreases for networks with
increasing anisotropy strength, i.e., for decreasing values of
$\sigma$. In other words, the robustness of anisotropic systems gets
reduced, compared to isotropic networks, as a result of the lack of
connections in the vertical direction, which rapidly get lost,
producing the fragmentation of the GC. This effect appears to be gradual
when increasing the anisotropy levels (see Fig.~\ref{gradual-sigma}
in the SI section). However, we expect that $l_c \to 0$ as $\sigma$ reaches
a given minimum value, below which a GC that spans the whole system cannot
be formed, prior to the localized attack, due to the insufficient amount
of vertical links, i.e., being below the percolation threshold. In addition,
and similar to isotropic networks, the critical linear size of attack does
not depend on $L$, for large system sizes. We can associate the differences
that the $L = 50$ curve presents, in Fig.~\ref{lc-vs-alpha} (d), with respect
to the results with larger $L$, to finite size effects, since networks have a
large characteristic link length compared to $L$ ($\zeta = 10$) and the
deviations appear more for large values of the attack size $l$ too. Furthermore,
it is plausible that the critical behavior of the system (isotropic or
anisotropic) is governed only by the adimensional fraction between lengths
$l/\zeta$. Indeed, our scaling results (see Fig.~\ref{lc-over-zeta} in the
SI section), suggest that such a scaling is a good approximation. However, more
extensive computations and better statistics should be obtained in order to
prove this assumption.

\begin{figure}[h]
  \subfloat{\begin{overpic}[width=7.0cm,height=5.0cm,angle=0]{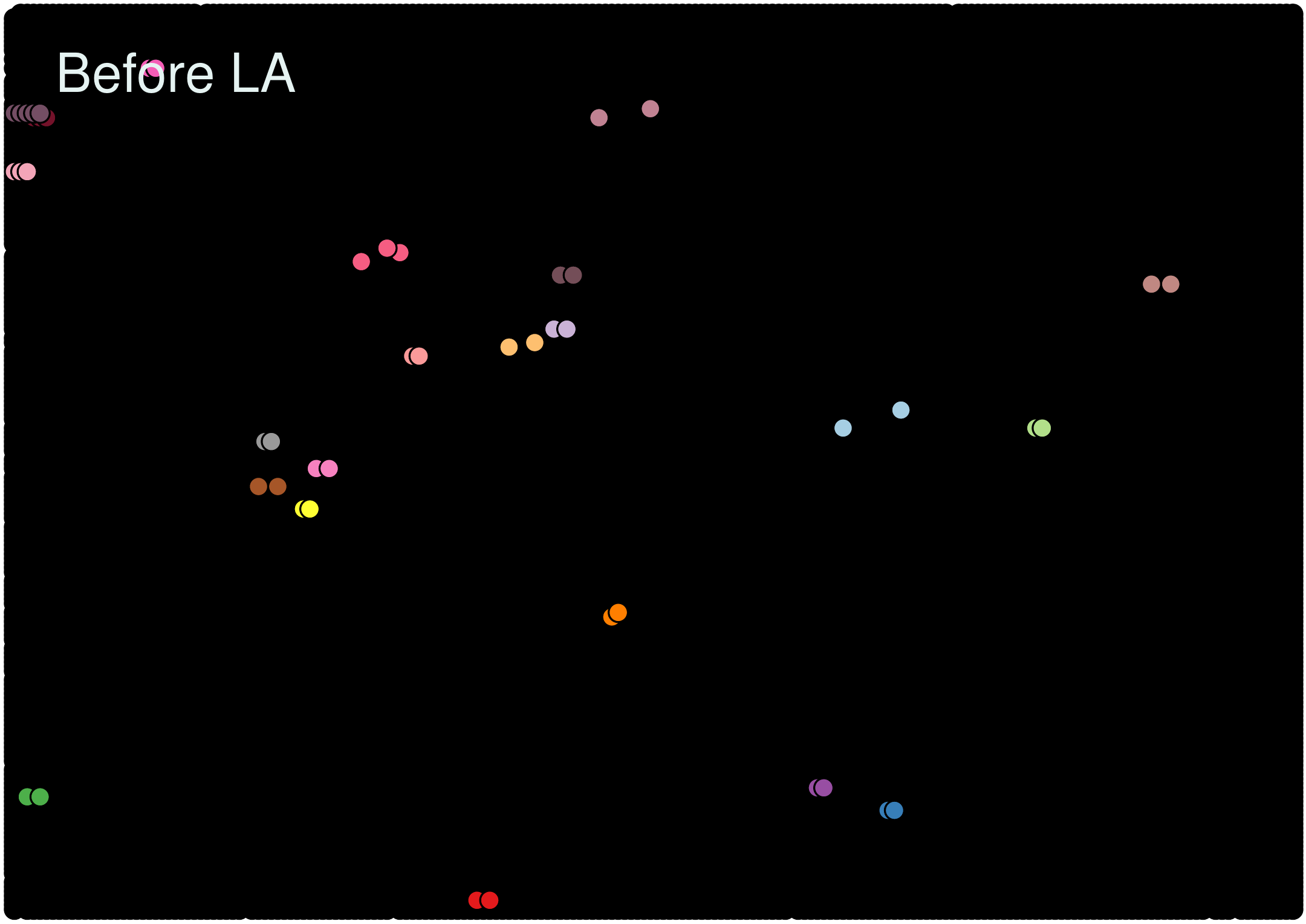}
      \put(67,63){\color{white}\bf{(a)} $\sigma = 0.25$}
  \end{overpic}}
  \hspace{0.1cm}
  \subfloat{\begin{overpic}[width=7.0cm,height=5.0cm,angle=0]{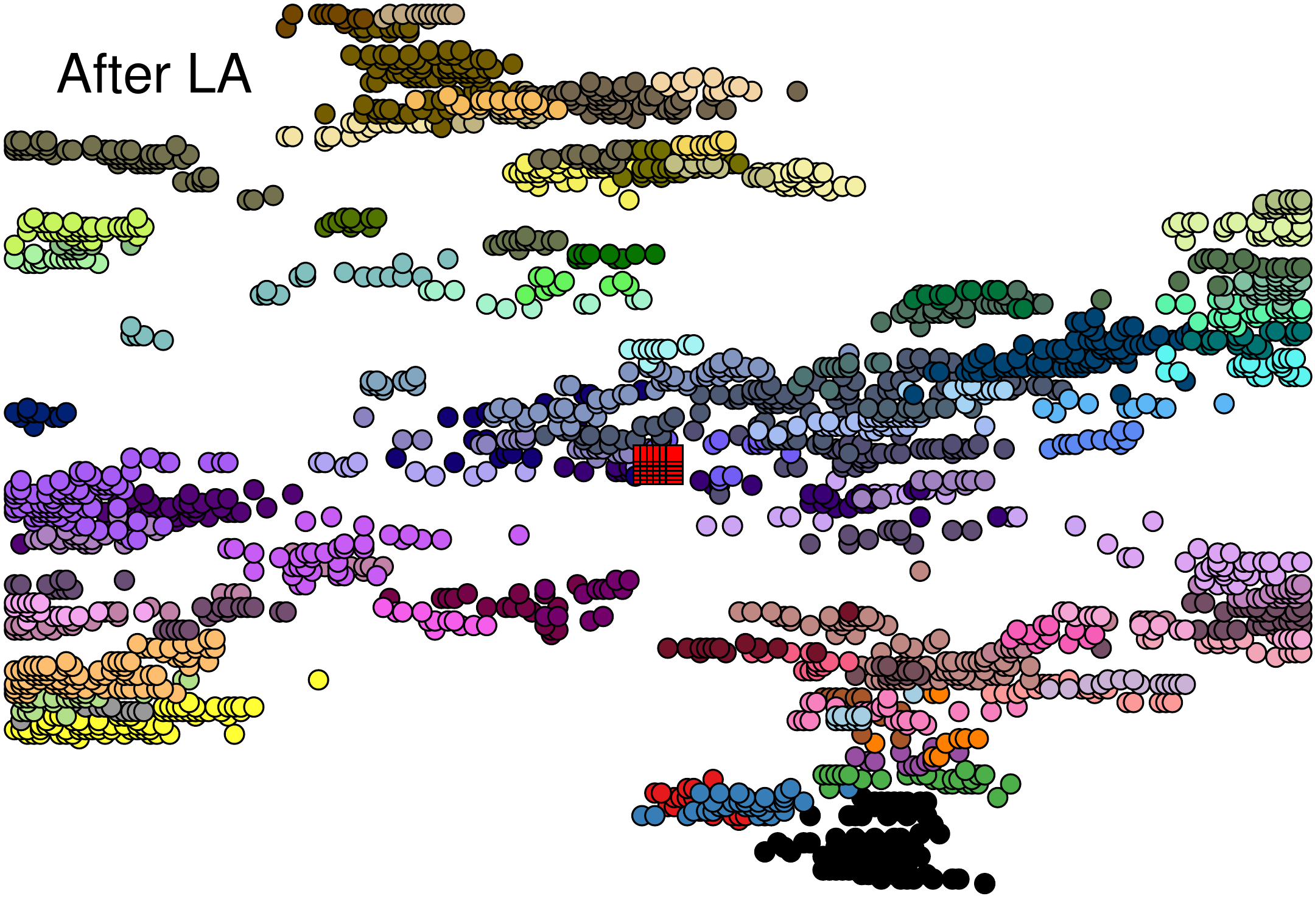}
    \put(67,63){\bf{(b)} $\sigma = 0.25$}
  \end{overpic}} \\
  \subfloat{\begin{overpic}[width=7.0cm,height=5.0cm,angle=0]{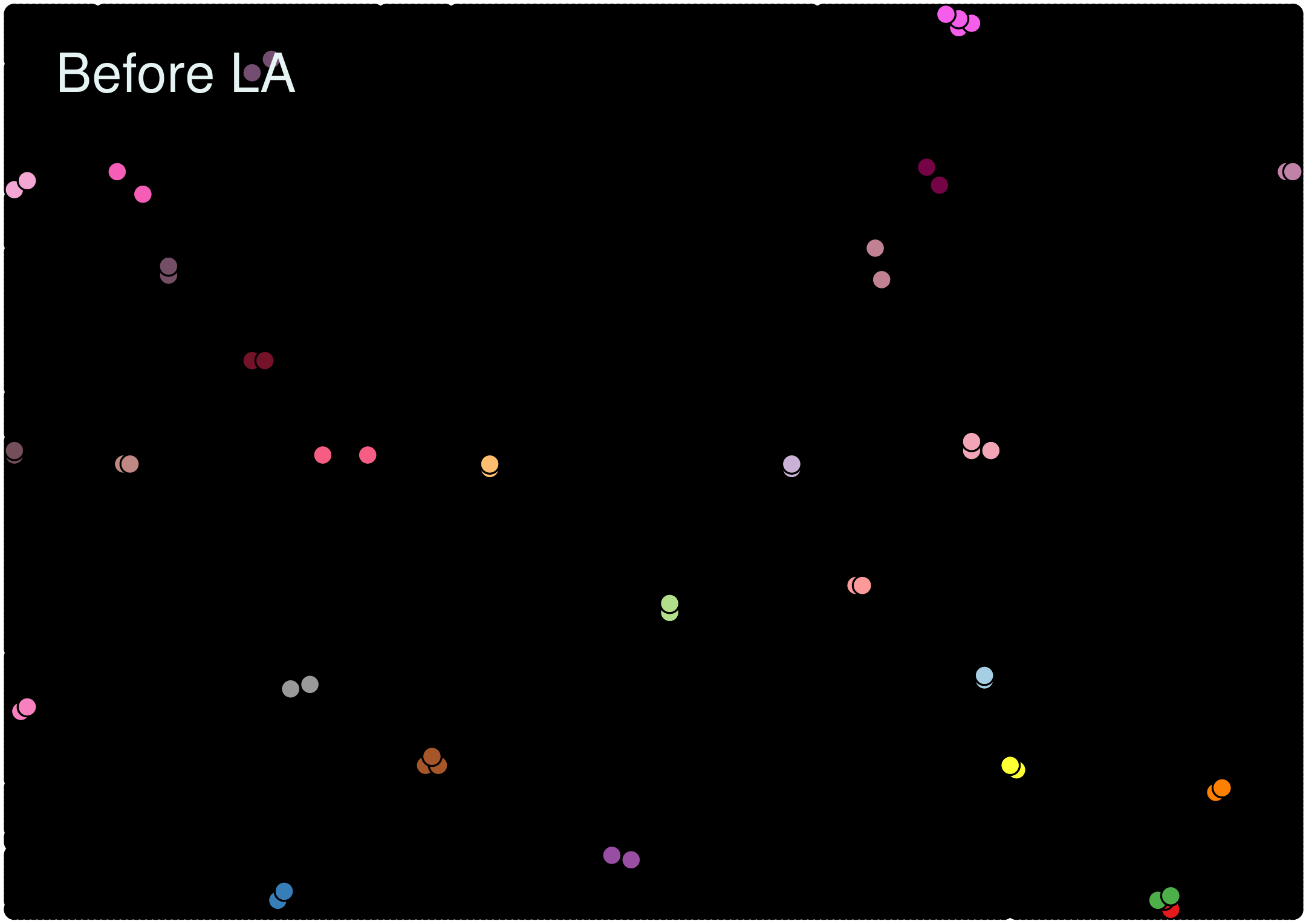}
      \put(75,4){\color{white}\bf{(c)} $\sigma = 1$}
  \end{overpic}}
  \hspace{0.1cm}
  \subfloat{\begin{overpic}[width=7.0cm,height=5.0cm,angle=0]{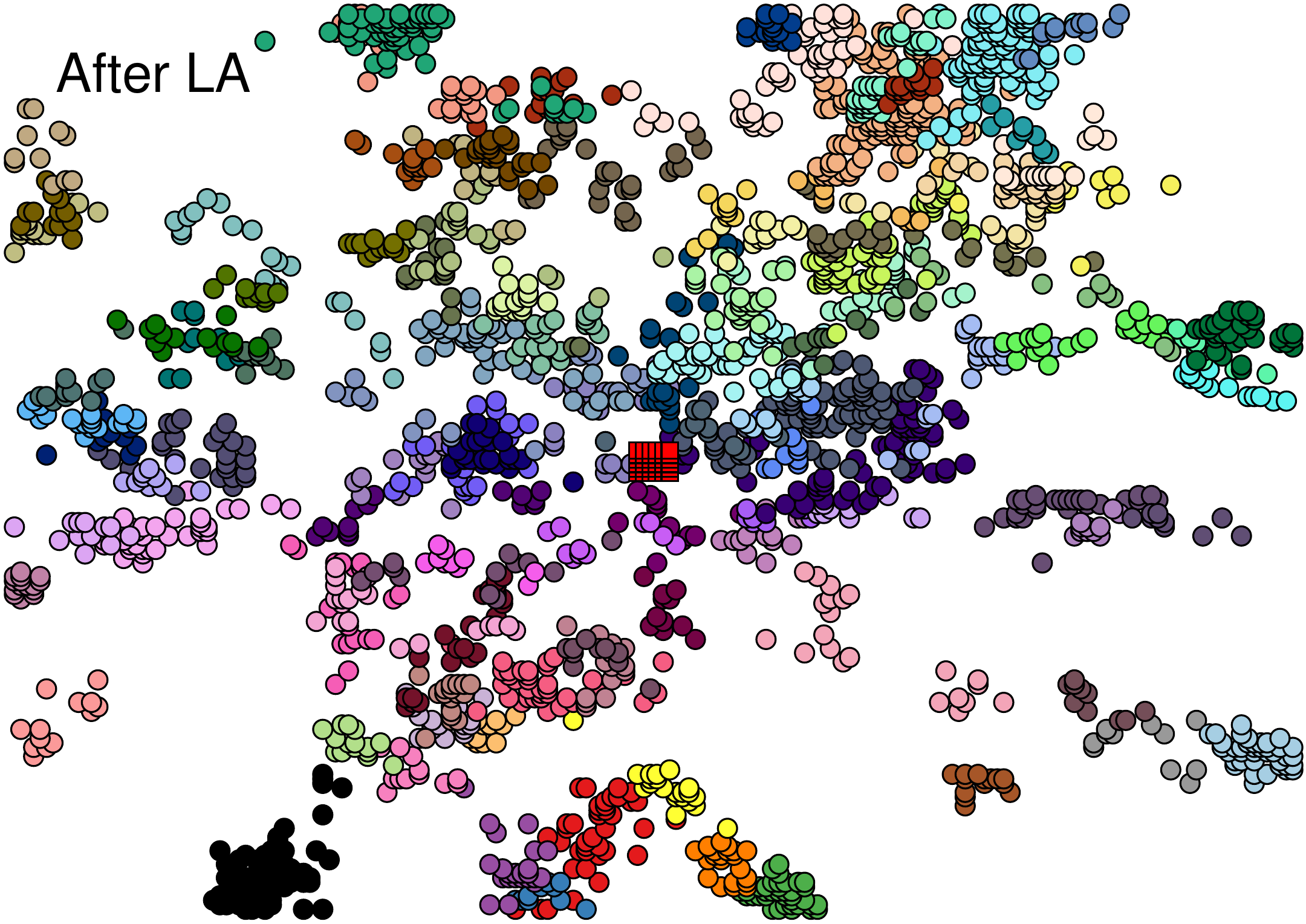}
      \put(75,4){\bf{(d)} $\sigma = 1$}
    \end{overpic}}
  \caption{Visualization of clusters before (left figures) and after
    (right figures) the localized attack (LA). In (a) and (b) we show
    an anisotropic network with $\sigma = 0.25$. From a GC that spans
    a huge part of the network (black region in (a)), we see that,
    after the LA, clusters retain some of the characteristic
    anisotropy, as they are rather stretched in the preferred
    horizontal direction. In (c) and (d) we present a more isotropic
    system, with $\sigma = 1$, in which the finite clusters, after the
    LA, do not have a defined shape. Note that red squares at the
    center of the networks in (b) and (d) represent the LA, of a square
    of linear size $l = 6$, while different clusters are depicted with
    circles of different colors. In order to simplify the display, in
    (b) and (d) we only show clusters with sizes between 10 and 100 nodes.
    The remaining parameters of the networks are $L = 200$, $\zeta = 3$,
    and $\alpha = 0.25$. Clusters shown are the result of a single
    realization of networks and the corresponding attacks.}
  \label{clusters}
\end{figure}

Lastly we analyze, at the end of the cascade process, all the
connected clusters of nodes excluding the giant component, defined as
the cluster touching all four edges of the system. First, in order to
visualize the spatial distribution and orientation of these clusters,
we plot them in Fig.~\ref{clusters}.
Figs.~\ref{clusters} (a) and (c), for $\sigma = 0.25$ and $\sigma =
1$, respectively, show that the GC (in black) spans almost the entire
network, {\it before} the localized attack (LA). On the other hand, at
the {\it end} of the process triggered by the attack, the GC becomes
fragmented and finite clusters appear, as seen in Figs.~\ref{clusters}
(b) and (d). Note that in the latter figures we only considered
clusters with sizes between 10 and 100 nodes, to clarify the image. We
observe that the shape of the finite clusters change with the
anisotropy of the system, and they can be located anywhere inside the
network. For $\sigma = 0.25$, clusters are mostly horizontally
stretched, while for more isotropic networks ($\sigma = 1$), the
clusters lose shape, as links can remain intact in many directions.

\begin{figure}[h]
  \subfloat{\begin{overpic}[width=7.0cm,height=5.0cm,angle=0]{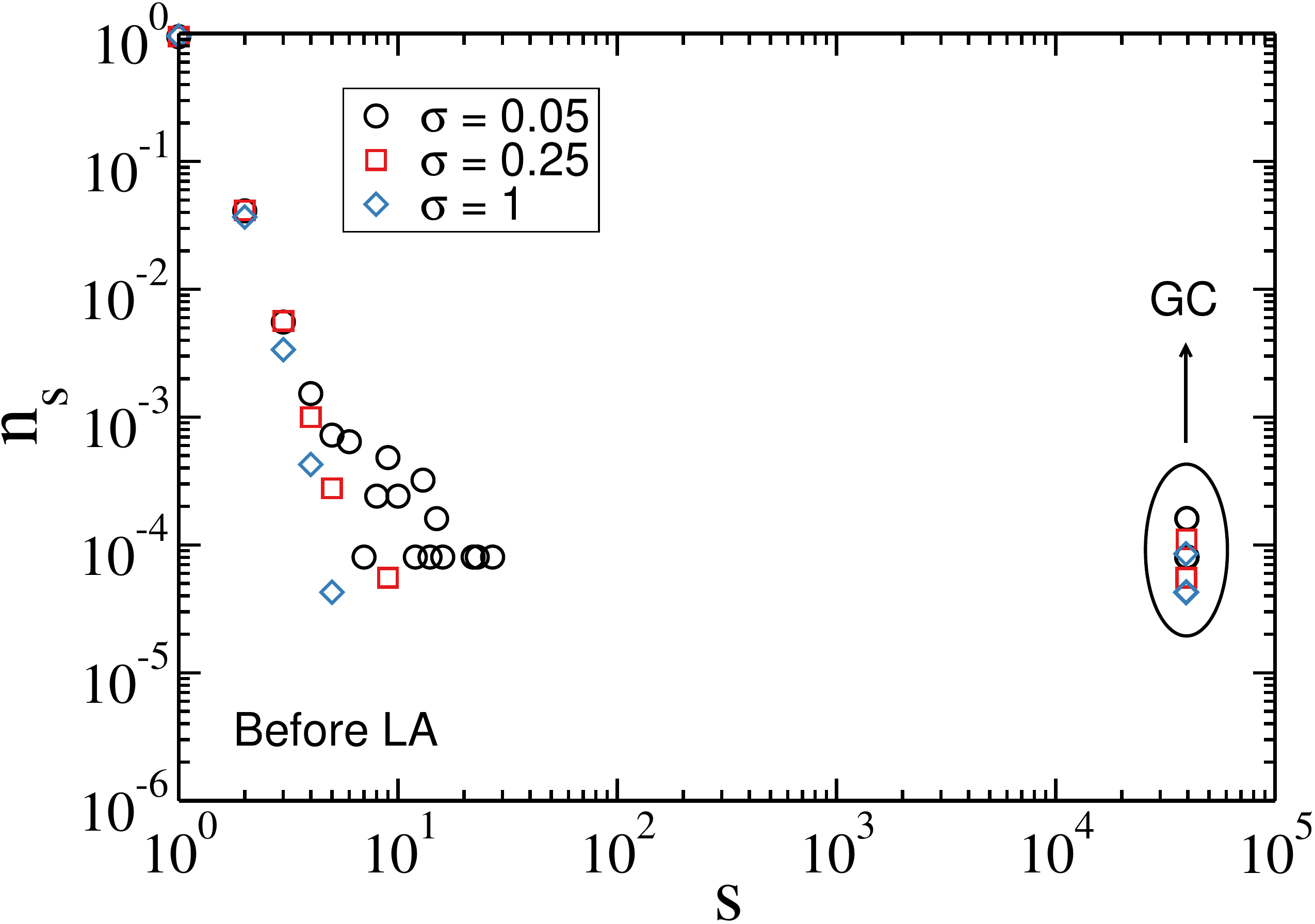}
      \put(68,61){\bf{(a)} $\zeta = 3$}
  \end{overpic}}
  \subfloat{\begin{overpic}[width=7.0cm,height=5.0cm,angle=0]{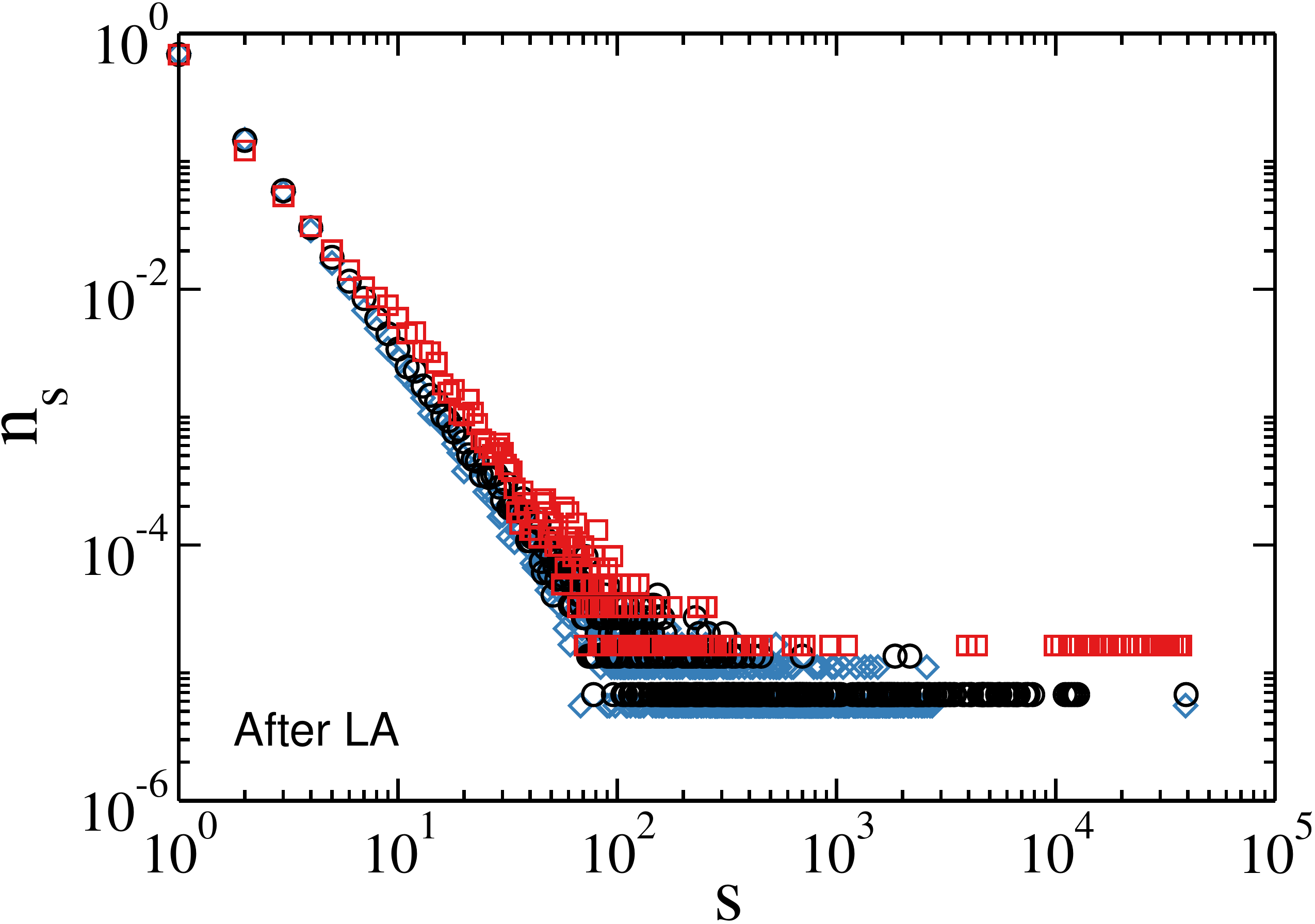}
    \put(68,61){\bf{(b)} $\zeta = 3$}
  \end{overpic}} \\
  \subfloat{\begin{overpic}[width=7.0cm,height=5.0cm,angle=0]{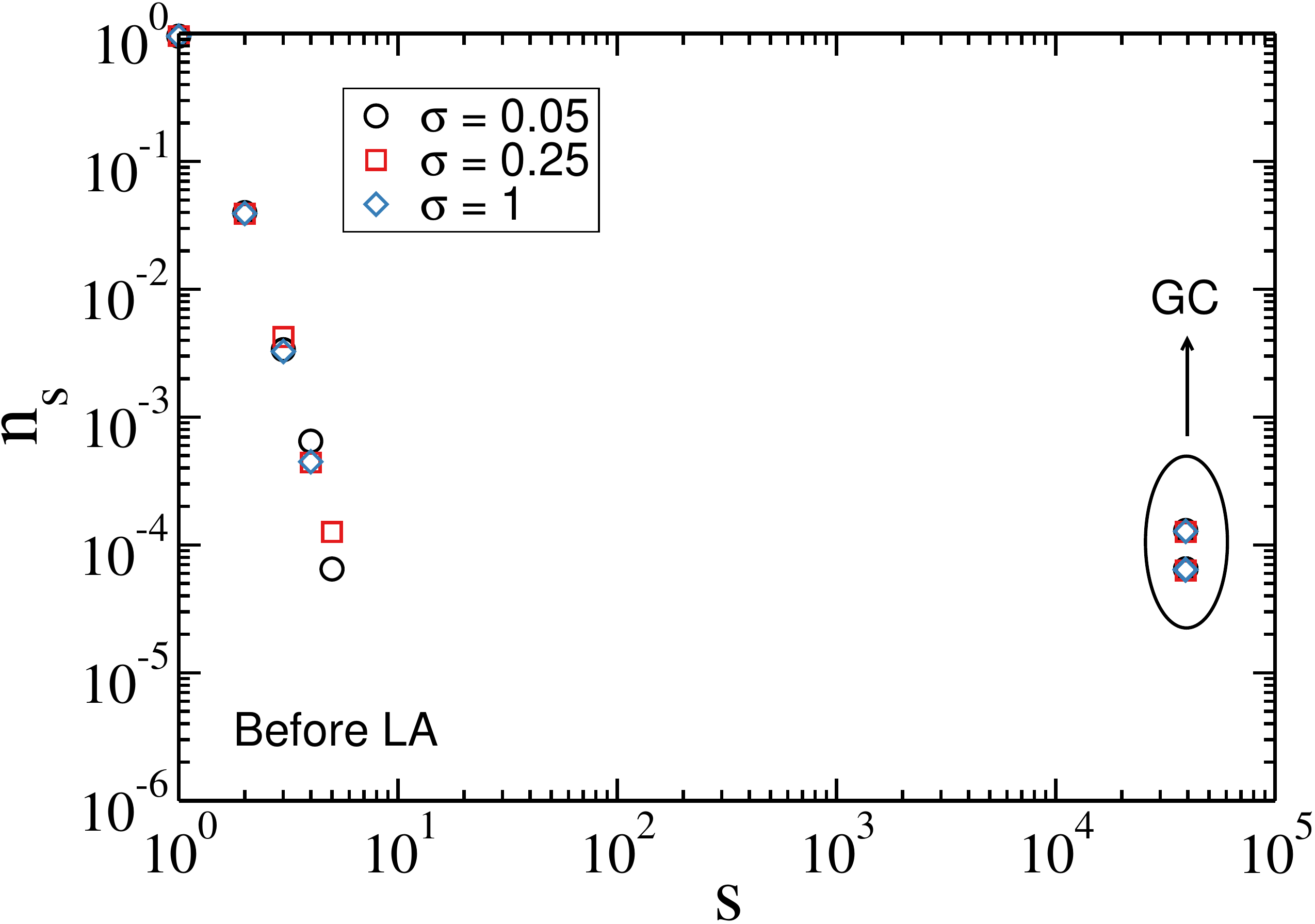}
      \put(68,61){\bf{(c)} $\zeta = 10$}
    \end{overpic}}
  \subfloat{\begin{overpic}[width=7.0cm,height=5.0cm,angle=0]{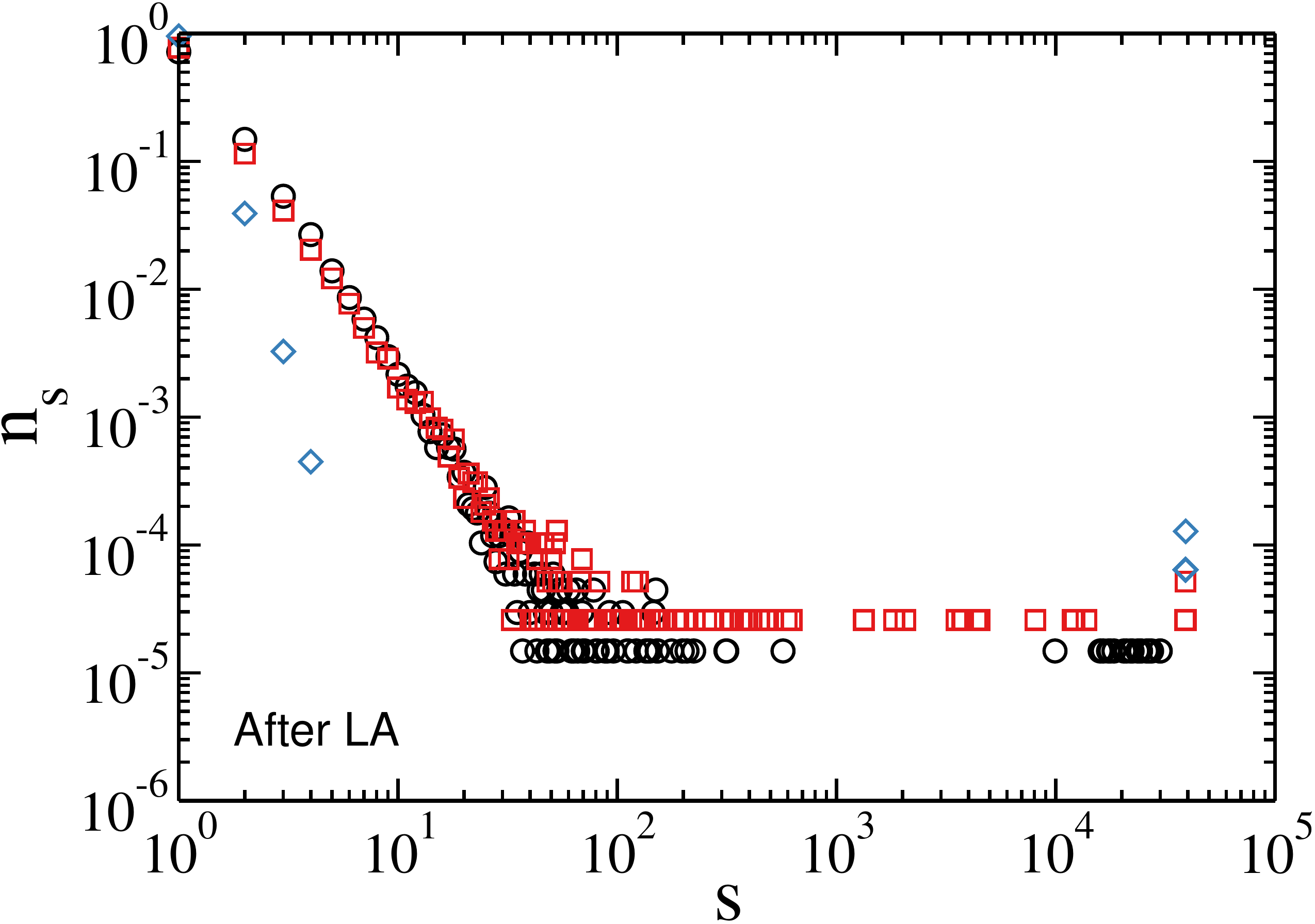}
      \put(68,61){\bf{(d)} $\zeta = 10$}
    \end{overpic}}
  \caption{Distribution of cluster sizes, $n_s$, before the localized
    attack (LA) of $l = 6$ (left figures) and after the cascade
    triggered by it stops (right figures). In (a) and (b) we show
    $n_s$ for $\zeta = 3$, while in (c) and (d) we show $n_s$ for
    $\zeta = 10$. The localized attack triggers a CF process that
    changes the original distribution of cluster sizes. In (d), for
    $\sigma = 1$, the distributions before and after the LA are
    practically the same, which indicates that the tolerance $\alpha =
    0.25$ is large enough to avoid the disintegration of the GC due to
    cascading failures. This is consistent with the critical linear
    size of attack, $l_c \approx 15$, found in Fig.~\ref{lc-vs-alpha}
    (d). When the CF develops in the whole network, a power-law
    behavior is observed in the final distribution of cluster sizes,
    i.e., $n_s \sim s^{-\tau}$, which suggests that the system is near
    the critical percolation state~\cite{coni-82}. These results correspond
    to networks of linear size $L = 200$, averaged over $N_{rea} = 70$
    realizations.} 
  \label{dist-bf-aft}
\end{figure}

Next, in Figs.~\ref{dist-bf-aft} (a) and (c) for $\zeta = 3$ and
$\zeta = 10$, respectively, we show the distribution $n_s \equiv
n_s(s)$ of cluster sizes, i.e., the probability of having a cluster of
size $s$, prior to the localized attack. The attack of $l = 6$, which is
above the critical size, triggers a cascade and affects the GC severely,
changing the distribution $n_s$, as shown in Figs.~\ref{dist-bf-aft} (b) and
(d), for $\alpha = 0.25$.
An interesting result that can be seen in Fig.~\ref{dist-bf-aft} is
that the final distribution of cluster sizes, $n_s$, behaves like a
power-law, i.e., $n_s \sim s^{-\tau}$, with $\tau > 0$. Observation of
the power-law in the cluster size distribution, at the end of the
cascade, is expected because the cascading failures drive the system
to the percolation critical point, at which the giant component
disappears. Note that in the Motter and Lai model~\cite{mott-02}, the
betweenness of nodes in a cluster is roughly proportional to the
square of its size. As soon as the giant component disappears, the
cascading failures stop because the betweenness of each node drops
below its maximal load.

\begin{figure}[h]
  \subfloat{\begin{overpic}[width=7.0cm,height=5.0cm,angle=0]{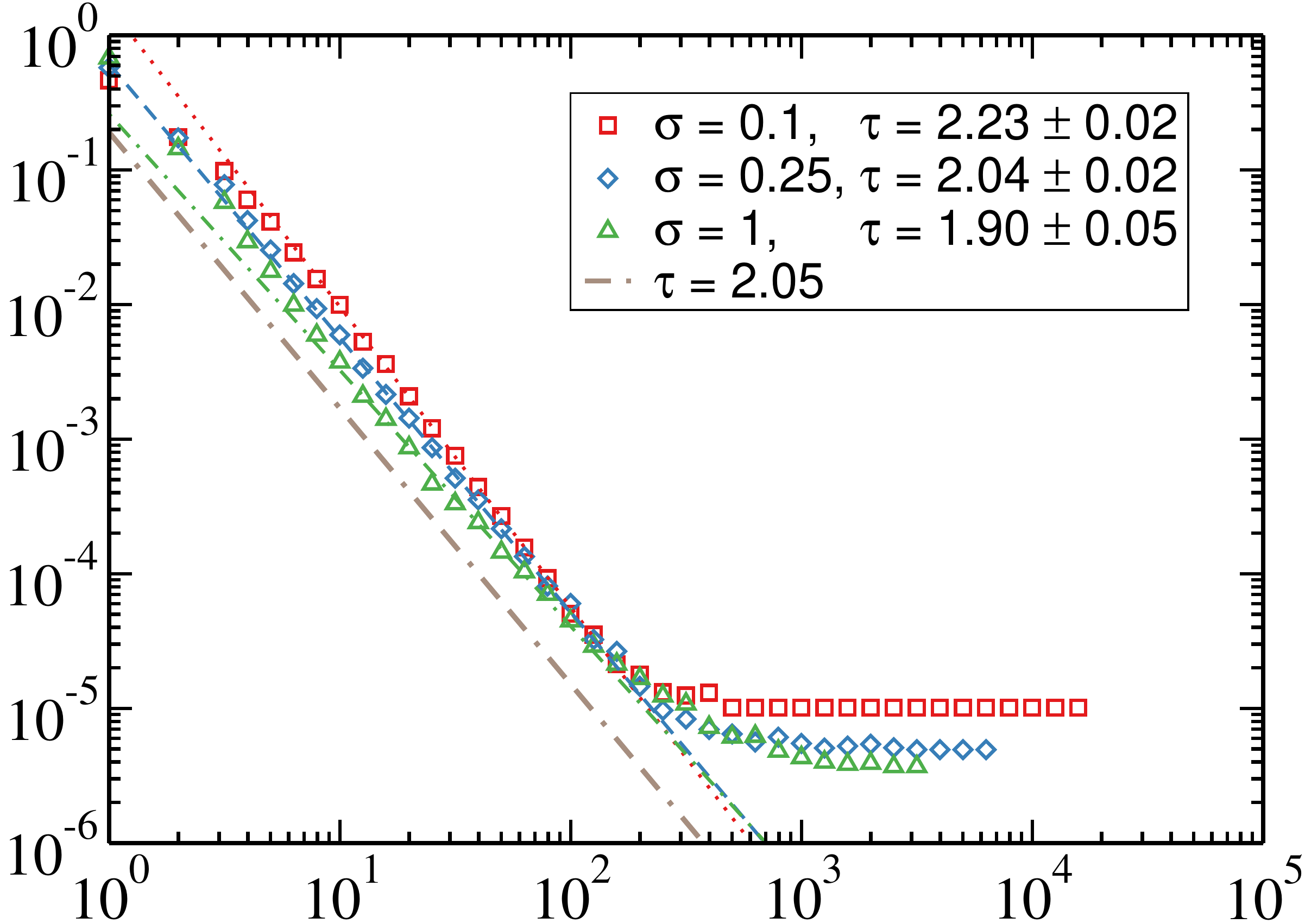}
      \put(12,12){\bf{(a)} $\zeta = 1$}
  \end{overpic}}
  \hspace{0.1cm}
  \subfloat{\begin{overpic}[width=7.0cm,height=5.0cm,angle=0]{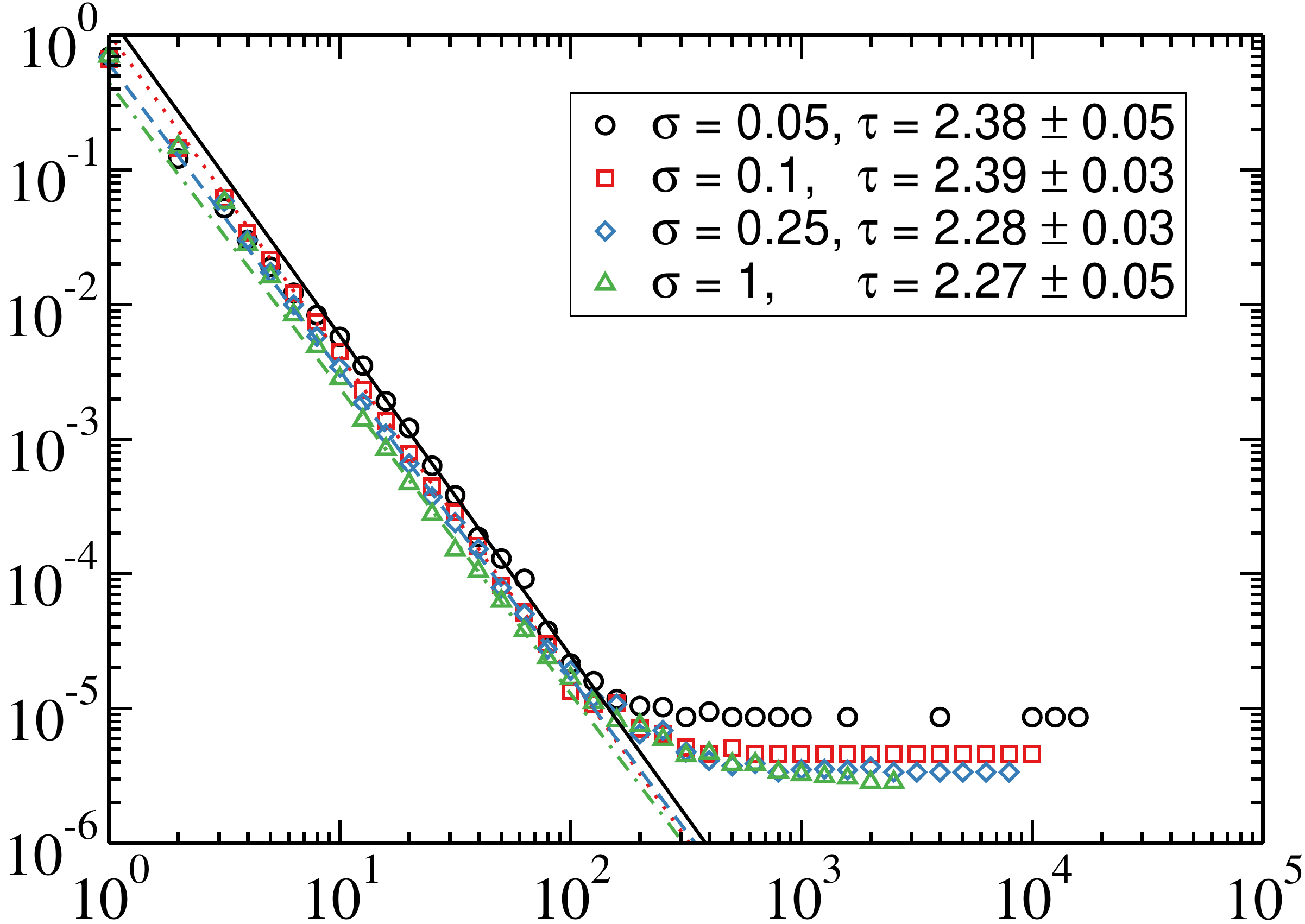}
      \put(12,12){\bf{(b)} $\zeta = 3$}
  \end{overpic}} \\
  \subfloat{\begin{overpic}[width=7.0cm,height=5.0cm,angle=0]{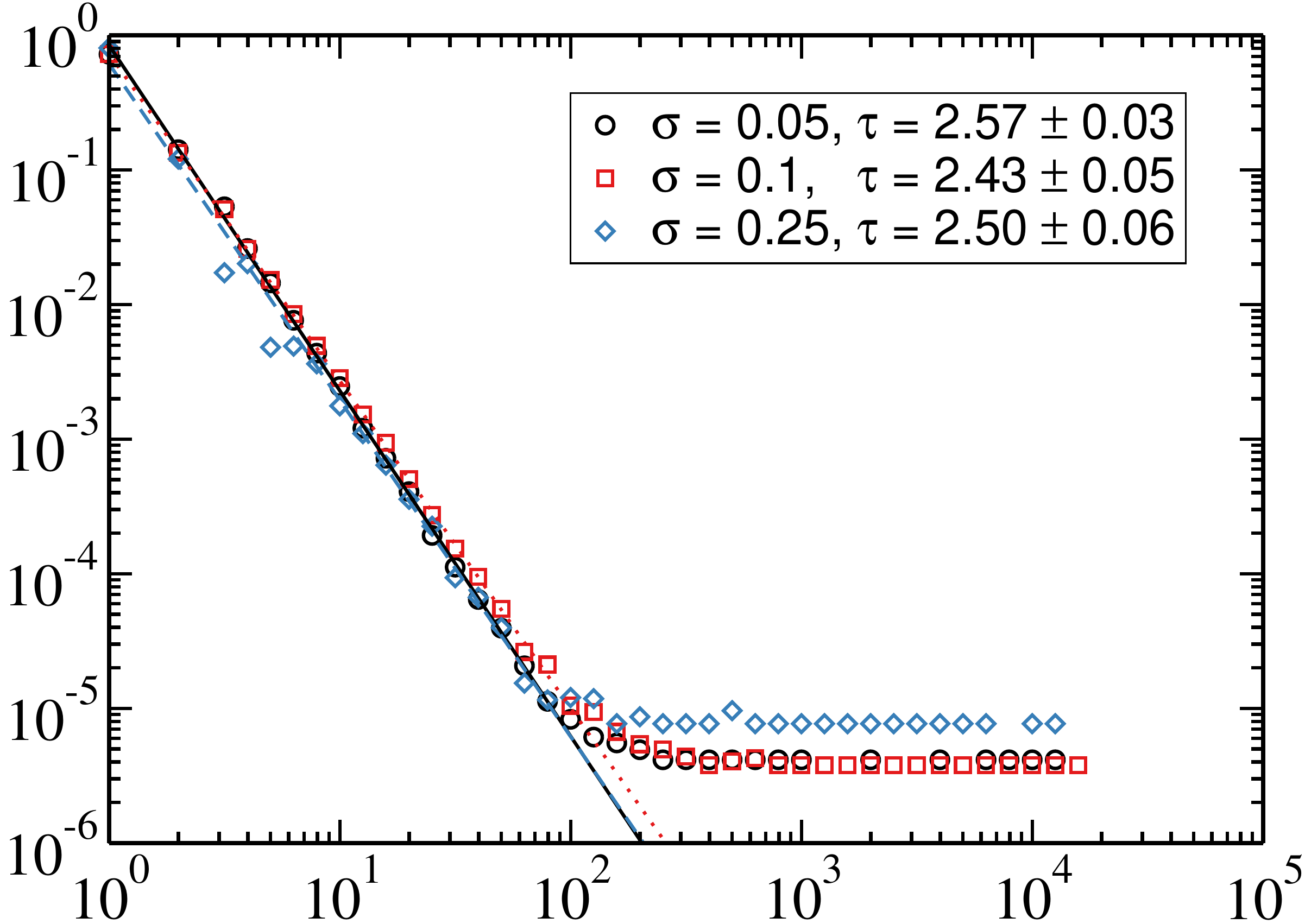}
      \put(12,12){\bf{(c)} $\zeta = 10$}
  \end{overpic}}
  \hspace{0.1cm}
  \subfloat{\begin{overpic}[width=7.0cm,height=5.0cm,angle=0]{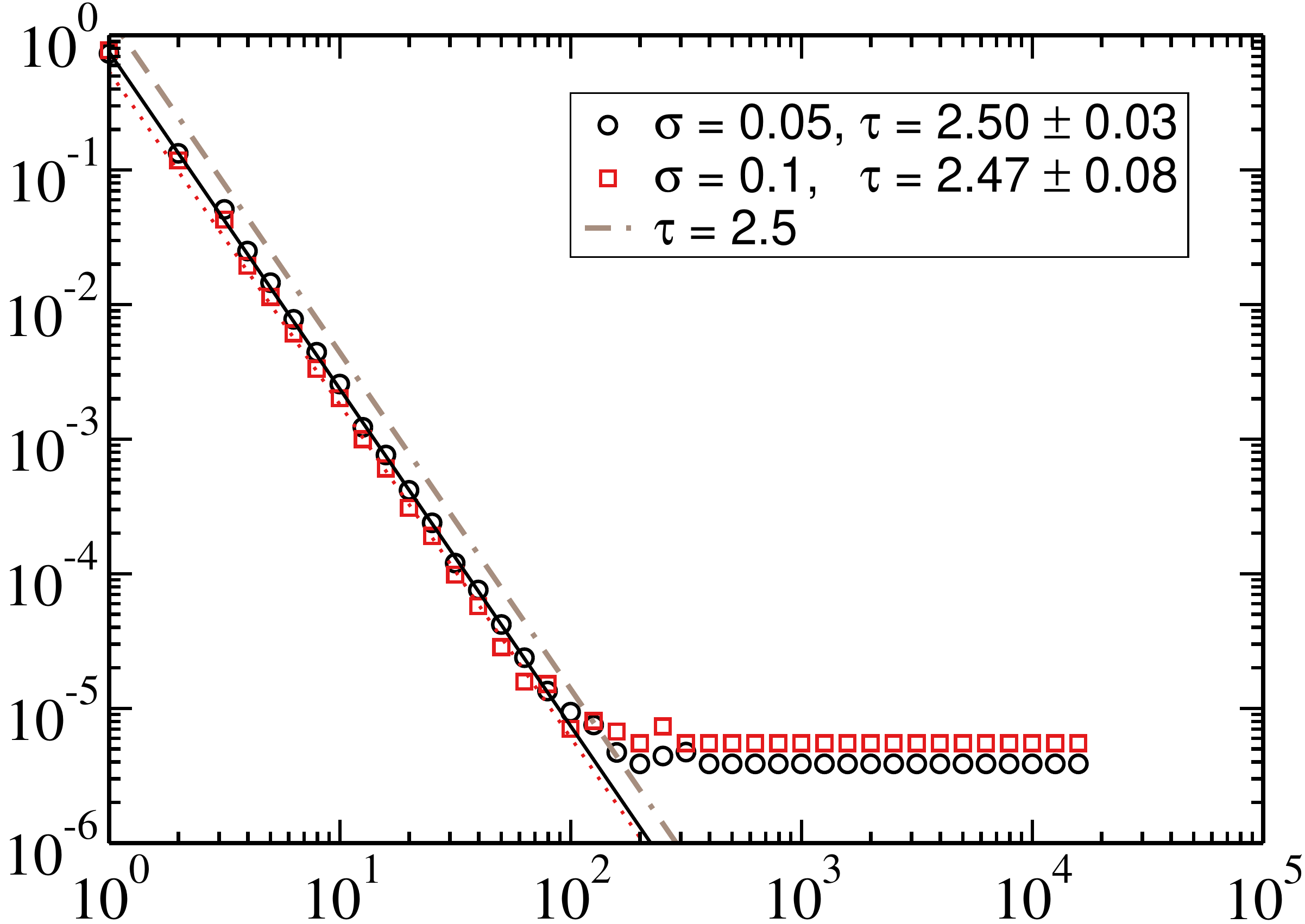}
      \put(12,12){\bf{(d)} $\zeta = 20$}
  \end{overpic}}
  \caption{Final distribution of cluster sizes $n_s$,
    corresponding to (a) $\zeta = 1$, (b) $\zeta = 3$, (c) $\zeta =
    10$, and (d) $\zeta = 20$. The distribution behaves partly (due to
    finite systems) like a power-law $n_s \sim s^{-\tau}$ (up to $s
    \sim 10^2$). Note that, for a strong spatial embedding - (a)
    $\zeta = 1$ -, the exponent $\tau$ is similar to that of critical
    percolation in a 2D-lattice, $\tau = 2.05$, while for large
    $\zeta$ of values 10 and 20, it approaches the value $\tau = 2.5$,
    corresponding to a random network with a Poisson degree
    distribution, i.e., mean field exponent. A logarithmic binning was
    applied to raw data, which allows a better perception of power-law
    behaviors. The remaining parameters of the networks are the same as in
    Fig.~\ref{dist-bf-aft}.}
  \label{ajuste-ns}
\end{figure}

Finally, we analyze the exponent $\tau$ of the power-law. When fitting the
initial trend of the data (up to $s \sim 10^2$) using logarithmic binning
we extract the exponents as shown in Fig.~\ref{ajuste-ns}.
For instance, in a network with strong spatial embedding ($\zeta = 1$ in
Fig.~\ref{ajuste-ns} (a)), the exponents are close to the value $\tau
= 2.05$, which corresponds to the known critical percolation exponent in
two-dimensional lattices~\cite{bunde-91, stauf-94}. Increasing the
characteristic link length $\zeta$, the spatial effects decline, and
the value of the power-law exponent tends to $\tau = 2.5$
(Fig.~\ref{ajuste-ns} (d)), which is characteristic of random networks
with a Poisson degree distribution or high dimensional
lattices~\cite{bunde-91, stauf-94} (with dimension $d \ge 6$). Note
that for $\zeta = 10$ and $\sigma = 1$ there is no cascade in the network,
due to its high tolerance levels. A similar situation occurs for
$\zeta = 20$ and $\sigma = 0.25, 1$, thus we did not include
these cases in Fig.~\ref{ajuste-ns} (d).

\section{Conclusions}

In this paper, we study the dynamic process of cascading failures
induced by overloads in both isotropic and anisotropic spatial
networks. We start the cascades from square shaped, localized attacks,
removing $l \times l$ nodes at the center of networks and analyze the
effects of the characteristic length $\zeta$ and the angular
dispersion $\sigma$ of links (which characterize the spatial embedding
and the anisotropy of the system, respectively) on the process.

First, we study the evolution and spatial distribution of the
failures, and find that anisotropy restricts the failures to spread
mostly along the preferential direction. This is a result of the lack
of connections in the orthogonal (vertical) direction, with respect to
the preferred (horizontal) direction, so the former become easily
overloaded.

Next, we find that there exists a critical damage linear size $l_c$,
above which the giant component of functional nodes collapses. This
critical size $l_c$ increases with the tolerance $\alpha$ and with the
strength of the spatial embedding, i.e., for small values of
$\zeta$. We find that anisotropy deteriorates the robustness of the
system, as the shortage of vertical links produces a weak connection
of the giant component along this direction, which become easily
overloaded and fail. In addition, these results, both for isotropic
and anisotropic networks, seem not to depend on the system size for large
enough systems. Thus, in the thermodynamic limit, a {\it zero} fraction
of localized failures yields a macroscopic transition. This is in marked
contrast to random failures, where a fraction of the system size (infinite
number of nodes in the thermodynamic limit) is needed to fail and cause
a system collapse.

\acknowledgments

IAP, CEL and LAB wish to thank to UNMdP (EXA 956/20), FONCyT (PICT
1422/2019) and CONICET, Argentina, for financial support. SH wishes to
thank the Israel Science Foundation, the Binational Israel-China
Science Foundation (Grant No. 3132/19), the BIU Center for Research in
Applied Cryptography and Cyber Security, NSF-BSF (Grant No. 2019740),
the EU H2020 project RISE (Project No. 821115), the EU H2020 DIT4TRAM,
and DTRA (Grant No. HDTRA-1-19-1-0016) for financial
support. D.V. thanks the PBC of the Council for Higher Education of
Israel for the Fellowship Grant. S.V.B acknowledges the partial
support of this research through the Dr. Bernard W. Gamson
Computational Science Center at Yeshiva College. This research was
supported by a grant from the United States-Israel Binational Science
Foundation (BSF), Jerusalem, Israel (Grant No. 2020255)."

\newpage
\section{Supplementary Information}
\begin{figure}[h]
  \subfloat{\begin{overpic}[width=7.0cm,height=5.0cm,angle=0]{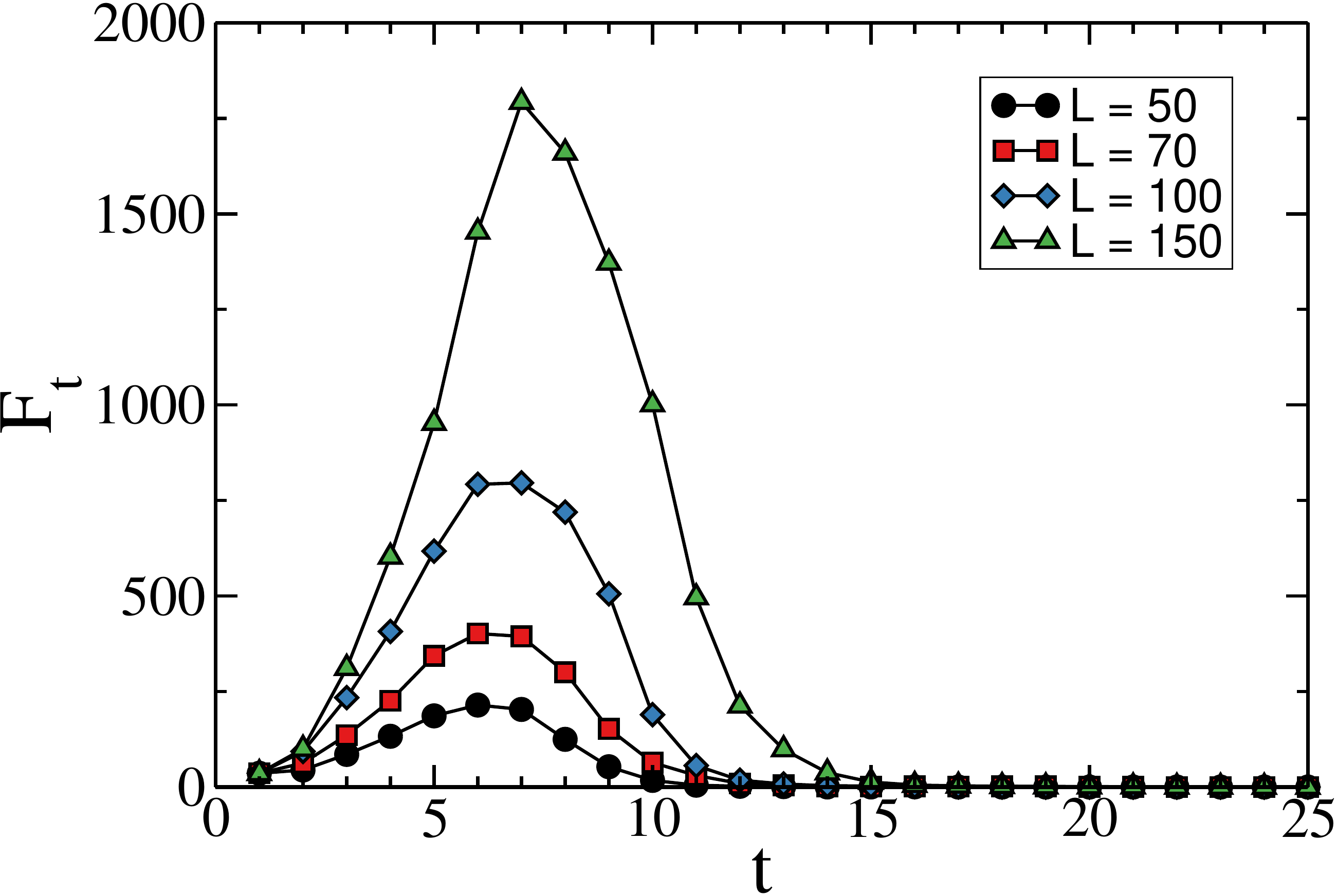}
      \put(0,0){\bf{(a)}}
    \end{overpic}}
  \subfloat{\begin{overpic}[width=7.0cm,height=5.0cm,angle=0]{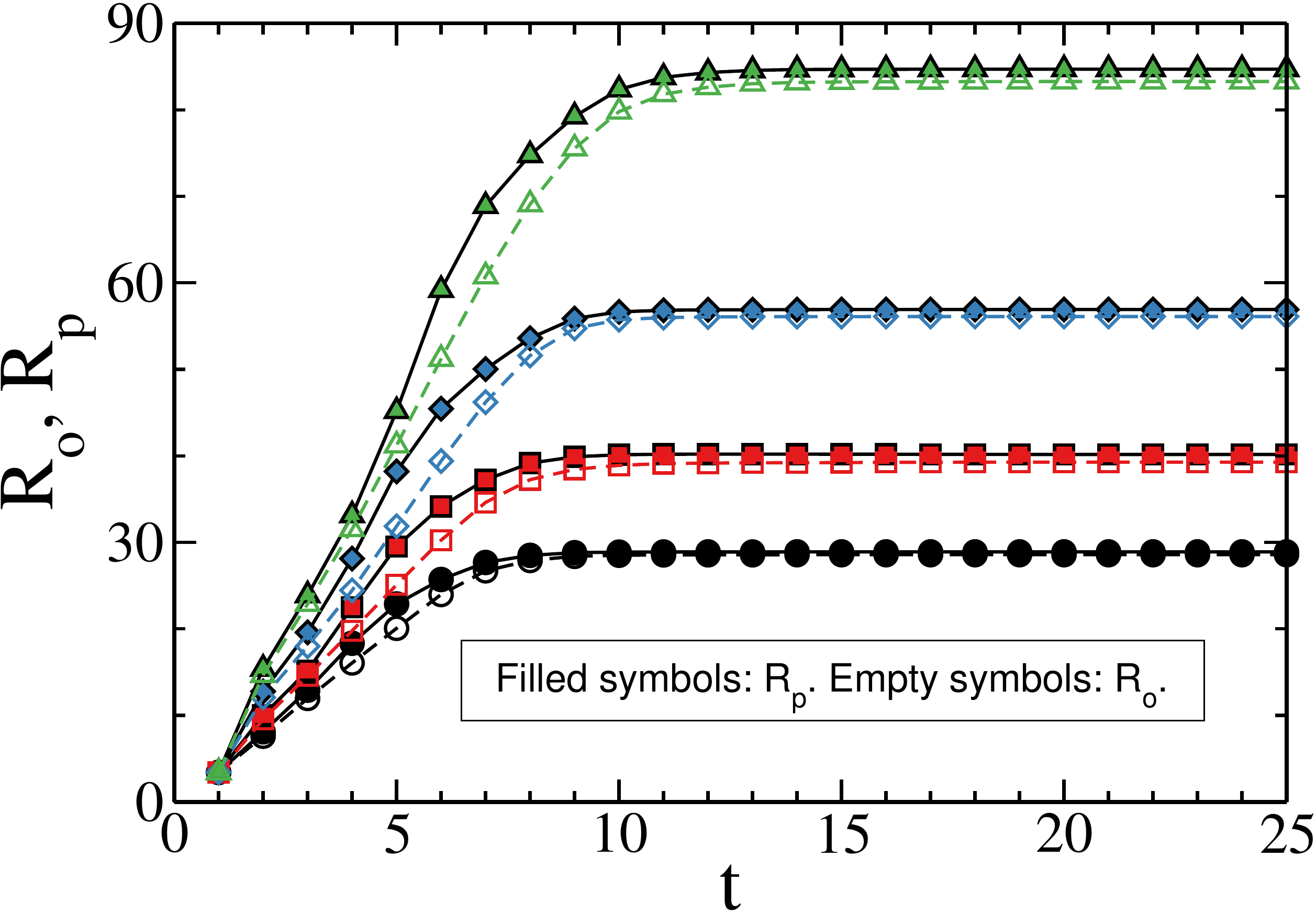}
      \put(0,0){\bf{(b)}}
    \end{overpic}}
    \caption{{\bf Consistency with isotropic model}. (a) Amount of
      failures $F_t$ at time $t$, for different values of the
      lattice-length $L$. (b) Estimation of the extent of the failures
      in the preferential ($\theta_p = 0$) and orthogonal ($\theta =
      \pi/2$) directions, $R_p$ and $R_o$, respectively, for the same
      values of $L$ than in (a). These results correspond to networks
      with strong spatial embedding ($\zeta = 1$) and low anisotropy
      ($\sigma = 1$), in which periodic boundary conditions were
      introduced, and which resemble the isotropic lattices used in
      Ref.~\cite{zhao-16}.  To be consistent with the measure of the
      magnitude $r_c(t)$ in~\cite{zhao-16}, the radius $R_p$ and $R_o$
      are multiplied by 2, to comprise approximately 95\% of all
      failures and, thus, represent the extent of the failures at time
      $t$. Our results are in well agreement with those obtained in
      the cited work. However, small differences appear due to the
      fact that our network is not a perfect lattice, and because of
      the different methodology for measuring the spatial distribution
      of the failures. Notice that the velocity of failure spreading
      (i.e., the slope of the radius), besides being approximately
      constant, is virtually the same both in the preferential and the
      orthogonal direction, given the low anisotropy of the network.
      The remaining parameters of the simulation are $\alpha = 0.25$,
      and $l = 6$. Results were averaged over $N_{rea} = 50$
      realizations.}\label{comp-iso-y-model}
\end{figure}
%
%


\begin{thebibliography}{18}%
\makeatletter
\providecommand \@ifxundefined [1]{%
 \@ifx{#1\undefined}
}%
\providecommand \@ifnum [1]{%
 \ifnum #1\expandafter \@firstoftwo
 \else \expandafter \@secondoftwo
 \fi
}%
\providecommand \@ifx [1]{%
 \ifx #1\expandafter \@firstoftwo
 \else \expandafter \@secondoftwo
 \fi
}%
\providecommand \natexlab [1]{#1}%
\providecommand \enquote  [1]{``#1''}%
\providecommand \bibnamefont  [1]{#1}%
\providecommand \bibfnamefont [1]{#1}%
\providecommand \citenamefont [1]{#1}%
\providecommand \href@noop [0]{\@secondoftwo}%
\providecommand \href [0]{\begingroup \@sanitize@url \@href}%
\providecommand \@href[1]{\@@startlink{#1}\@@href}%
\providecommand \@@href[1]{\endgroup#1\@@endlink}%
\providecommand \@sanitize@url [0]{\catcode `\\12\catcode `\$12\catcode
  `\&12\catcode `\#12\catcode `\^12\catcode `\_12\catcode `\%12\relax}%
\providecommand \@@startlink[1]{}%
\providecommand \@@endlink[0]{}%
\providecommand \url  [0]{\begingroup\@sanitize@url \@url }%
\providecommand \@url [1]{\endgroup\@href {#1}{\urlprefix }}%
\providecommand \urlprefix  [0]{URL }%
\providecommand \Eprint [0]{\href }%
\providecommand \doibase [0]{http://dx.doi.org/}%
\providecommand \selectlanguage [0]{\@gobble}%
\providecommand \bibinfo  [0]{\@secondoftwo}%
\providecommand \bibfield  [0]{\@secondoftwo}%
\providecommand \translation [1]{[#1]}%
\providecommand \BibitemOpen [0]{}%
\providecommand \bibitemStop [0]{}%
\providecommand \bibitemNoStop [0]{.\EOS\space}%
\providecommand \EOS [0]{\spacefactor3000\relax}%
\providecommand \BibitemShut  [1]{\csname bibitem#1\endcsname}%
\let\auto@bib@innerbib\@empty
\bibitem [{\citenamefont {Motter}\ and\ \citenamefont {Lai}(2002)}]{mott-02}%
  \BibitemOpen
  \bibfield  {author} {\bibinfo {author} {\bibfnamefont {A.~E.}\ \bibnamefont
  {Motter}}\ and\ \bibinfo {author} {\bibfnamefont {Y.-C.}\ \bibnamefont
  {Lai}},\ }\href {\doibase 10.1103/PhysRevE.66.065102} {\bibfield  {journal}
  {\bibinfo  {journal} {Phys. Rev. E}\ }\textbf {\bibinfo {volume} {66}},\
  \bibinfo {pages} {065102} (\bibinfo {year} {2002})}\BibitemShut {NoStop}%
\bibitem [{\citenamefont {Berezin}\ \emph {et~al.}(2015)\citenamefont
  {Berezin}, \citenamefont {Bashan}, \citenamefont {Danziger}, \citenamefont
  {Daqing},\ and\ \citenamefont {Havlin}}]{berez-15}%
  \BibitemOpen
  \bibfield  {author} {\bibinfo {author} {\bibfnamefont {Y.}~\bibnamefont
  {Berezin}}, \bibinfo {author} {\bibfnamefont {A.}~\bibnamefont {Bashan}},
  \bibinfo {author} {\bibfnamefont {M.}~\bibnamefont {Danziger}}, \bibinfo
  {author} {\bibfnamefont {L.}~\bibnamefont {Daqing}}, \ and\ \bibinfo {author}
  {\bibfnamefont {S.}~\bibnamefont {Havlin}},\ }\href {\doibase
  10.1038/srep08934} {\bibfield  {journal} {\bibinfo  {journal} {Scientific
  reports}\ }\textbf {\bibinfo {volume} {5}},\ \bibinfo {pages} {8934}
  (\bibinfo {year} {2015})}\BibitemShut {NoStop}%
\bibitem [{\citenamefont {Newman}(2010)}]{new-10}%
  \BibitemOpen
  \bibfield  {author} {\bibinfo {author} {\bibfnamefont {M.~E.~J.}\
  \bibnamefont {Newman}},\ }\href {\doibase
  10.1093/acprof:oso/9780199206650.001.0001} {\emph {\bibinfo {title}
  {Networks: An Introduction}}}\ (\bibinfo  {publisher} {Oxford University
  Press},\ \bibinfo {year} {2010})\BibitemShut {NoStop}%
\bibitem [{\citenamefont {Bunde}\ and\ \citenamefont
  {Havlin}(1991)}]{bunde-91}%
  \BibitemOpen
  \bibfield  {author} {\bibinfo {author} {\bibfnamefont {A.}~\bibnamefont
  {Bunde}}\ and\ \bibinfo {author} {\bibfnamefont {S.}~\bibnamefont {Havlin}},\
  }\href@noop {} {\emph {\bibinfo {title} {Fractals and disordered systems}}}\
  (\bibinfo  {publisher} {Springer-Verlag New York, Inc.},\ \bibinfo {year}
  {1991})\BibitemShut {NoStop}%
\bibitem [{\citenamefont {Kornbluth}\ \emph {et~al.}(2018)\citenamefont
  {Kornbluth}, \citenamefont {Barach}, \citenamefont {Tuchman}, \citenamefont
  {Kadish}, \citenamefont {Cwilich},\ and\ \citenamefont {Buldyrev}}]{korn-18}%
  \BibitemOpen
  \bibfield  {author} {\bibinfo {author} {\bibfnamefont {Y.}~\bibnamefont
  {Kornbluth}}, \bibinfo {author} {\bibfnamefont {G.}~\bibnamefont {Barach}},
  \bibinfo {author} {\bibfnamefont {Y.}~\bibnamefont {Tuchman}}, \bibinfo
  {author} {\bibfnamefont {B.}~\bibnamefont {Kadish}}, \bibinfo {author}
  {\bibfnamefont {G.}~\bibnamefont {Cwilich}}, \ and\ \bibinfo {author}
  {\bibfnamefont {S.~V.}\ \bibnamefont {Buldyrev}},\ }\href {\doibase
  10.1103/PhysRevE.97.052309} {\bibfield  {journal} {\bibinfo  {journal} {Phys.
  Rev. E}\ }\textbf {\bibinfo {volume} {97}},\ \bibinfo {pages} {052309}
  (\bibinfo {year} {2018})}\BibitemShut {NoStop}%
\bibitem [{\citenamefont {Dobson}\ and\ \citenamefont {Lu}(1992)}]{dobs-92}%
  \BibitemOpen
  \bibfield  {author} {\bibinfo {author} {\bibfnamefont {I.}~\bibnamefont
  {Dobson}}\ and\ \bibinfo {author} {\bibfnamefont {L.}~\bibnamefont {Lu}},\
  }\href {\doibase 10.1109/81.250167} {\bibfield  {journal} {\bibinfo
  {journal} {IEEE Transactions on Circuits and Systems I: Fundamental Theory
  and Applications}\ }\textbf {\bibinfo {volume} {39}},\ \bibinfo {pages} {762}
  (\bibinfo {year} {1992})}\BibitemShut {NoStop}%
\bibitem [{\citenamefont {Dobson}\ \emph {et~al.}(2007)\citenamefont {Dobson},
  \citenamefont {Carreras}, \citenamefont {Lynch},\ and\ \citenamefont
  {Newman}}]{dobs-07}%
  \BibitemOpen
  \bibfield  {author} {\bibinfo {author} {\bibfnamefont {I.}~\bibnamefont
  {Dobson}}, \bibinfo {author} {\bibfnamefont {B.~A.}\ \bibnamefont
  {Carreras}}, \bibinfo {author} {\bibfnamefont {V.~E.}\ \bibnamefont {Lynch}},
  \ and\ \bibinfo {author} {\bibfnamefont {D.~E.}\ \bibnamefont {Newman}},\
  }\href {\doibase 10.1063/1.2737822} {\bibfield  {journal} {\bibinfo
  {journal} {Chaos: An Interdisciplinary Journal of Nonlinear Science}\
  }\textbf {\bibinfo {volume} {17}},\ \bibinfo {pages} {026103} (\bibinfo
  {year} {2007})}\BibitemShut {NoStop}%
\bibitem [{\citenamefont {Carreras}\ \emph {et~al.}(2016)\citenamefont
  {Carreras}, \citenamefont {Newman},\ and\ \citenamefont {Dobson}}]{carr-16}%
  \BibitemOpen
  \bibfield  {author} {\bibinfo {author} {\bibfnamefont {B.~A.}\ \bibnamefont
  {Carreras}}, \bibinfo {author} {\bibfnamefont {D.~E.}\ \bibnamefont
  {Newman}}, \ and\ \bibinfo {author} {\bibfnamefont {I.}~\bibnamefont
  {Dobson}},\ }\href {\doibase 10.1109/TPWRS.2015.2510627} {\bibfield
  {journal} {\bibinfo  {journal} {IEEE Transactions on Power Systems}\ }\textbf
  {\bibinfo {volume} {31}},\ \bibinfo {pages} {4406} (\bibinfo {year}
  {2016})}\BibitemShut {NoStop}%
\bibitem [{\citenamefont {Zhao}\ \emph {et~al.}(2016)\citenamefont {Zhao},
  \citenamefont {Li}, \citenamefont {Sanhedrai}, \citenamefont {Cohen},\ and\
  \citenamefont {Havlin}}]{zhao-16}%
  \BibitemOpen
  \bibfield  {author} {\bibinfo {author} {\bibfnamefont {J.}~\bibnamefont
  {Zhao}}, \bibinfo {author} {\bibfnamefont {D.}~\bibnamefont {Li}}, \bibinfo
  {author} {\bibfnamefont {H.}~\bibnamefont {Sanhedrai}}, \bibinfo {author}
  {\bibfnamefont {R.}~\bibnamefont {Cohen}}, \ and\ \bibinfo {author}
  {\bibfnamefont {S.}~\bibnamefont {Havlin}},\ }\href@noop {} {\bibfield
  {journal} {\bibinfo  {journal} {Nature communications}\ }\textbf {\bibinfo
  {volume} {7}},\ \bibinfo {pages} {1 } (\bibinfo {year} {2016})}\BibitemShut
  {NoStop}%
\bibitem [{\citenamefont {Vaknin}\ \emph {et~al.}(2017)\citenamefont {Vaknin},
  \citenamefont {Danziger},\ and\ \citenamefont {Havlin}}]{vak-17}%
  \BibitemOpen
  \bibfield  {author} {\bibinfo {author} {\bibfnamefont {D.}~\bibnamefont
  {Vaknin}}, \bibinfo {author} {\bibfnamefont {M.~M.}\ \bibnamefont
  {Danziger}}, \ and\ \bibinfo {author} {\bibfnamefont {S.}~\bibnamefont
  {Havlin}},\ }\href {\doibase 10.1088/1367-2630/aa7b09} {\bibfield  {journal}
  {\bibinfo  {journal} {New Journal of Physics}\ }\textbf {\bibinfo {volume}
  {19}},\ \bibinfo {pages} {073037} (\bibinfo {year} {2017})}\BibitemShut
  {NoStop}%
\bibitem [{\citenamefont {Waxman}(1988)}]{wax-88}%
  \BibitemOpen
  \bibfield  {author} {\bibinfo {author} {\bibfnamefont {B.~M.}\ \bibnamefont
  {Waxman}},\ }\href@noop {} {\bibfield  {journal} {\bibinfo  {journal} {IEEE
  J. Sel. Areas Commun.}\ }\textbf {\bibinfo {volume} {6}},\ \bibinfo {pages}
  {1617} (\bibinfo {year} {1988})}\BibitemShut {NoStop}%
\bibitem [{\citenamefont {Daqing}\ \emph {et~al.}(2011)\citenamefont {Daqing},
  \citenamefont {Kosmidis}, \citenamefont {Bunde},\ and\ \citenamefont
  {Havlin}}]{daq-11}%
  \BibitemOpen
  \bibfield  {author} {\bibinfo {author} {\bibfnamefont {L.}~\bibnamefont
  {Daqing}}, \bibinfo {author} {\bibfnamefont {K.}~\bibnamefont {Kosmidis}},
  \bibinfo {author} {\bibfnamefont {A.}~\bibnamefont {Bunde}}, \ and\ \bibinfo
  {author} {\bibfnamefont {S.}~\bibnamefont {Havlin}},\ }\href {\doibase
  10.1038/nphys1932} {\bibfield  {journal} {\bibinfo  {journal} {Nature
  Physics}\ }\textbf {\bibinfo {volume} {7}},\ \bibinfo {pages} {481} (\bibinfo
  {year} {2011})}\BibitemShut {NoStop}%
\bibitem [{\citenamefont {{National Land Information Division, National Spatial
  Planning and Regional Policy Bureau, MILT of Japan}}(2012)}]{2012japan}%
  \BibitemOpen
  \bibfield  {author} {\bibinfo {author} {\bibnamefont {{National Land
  Information Division, National Spatial Planning and Regional Policy Bureau,
  MILT of Japan}}},\ }\href@noop {} {\bibfield  {journal} {\bibinfo  {journal}
  {National railway data}\ } (\bibinfo {year} {2012})}\BibitemShut {NoStop}%
\bibitem [{\citenamefont {Danziger}\ \emph {et~al.}(2016)\citenamefont
  {Danziger}, \citenamefont {Shekhtman}, \citenamefont {Berezin},\ and\
  \citenamefont {Havlin}}]{dan-16}%
  \BibitemOpen
  \bibfield  {author} {\bibinfo {author} {\bibfnamefont {M.~M.}\ \bibnamefont
  {Danziger}}, \bibinfo {author} {\bibfnamefont {L.~M.}\ \bibnamefont
  {Shekhtman}}, \bibinfo {author} {\bibfnamefont {Y.}~\bibnamefont {Berezin}},
  \ and\ \bibinfo {author} {\bibfnamefont {S.}~\bibnamefont {Havlin}},\ }\href
  {\doibase 10.1209/0295-5075/115/36002} {\bibfield  {journal} {\bibinfo
  {journal} {{EPL} (Europhysics Letters)}\ }\textbf {\bibinfo {volume} {115}},\
  \bibinfo {pages} {36002} (\bibinfo {year} {2016})}\BibitemShut {NoStop}%
\bibitem [{\citenamefont {Havlin}\ \emph {et~al.}(2005)\citenamefont {Havlin},
  \citenamefont {Braunstein}, \citenamefont {Buldyrev}, \citenamefont {Cohen},
  \citenamefont {Kalisky}, \citenamefont {Sreenivasan},\ and\ \citenamefont
  {{Eugene Stanley}}}]{havl-05}%
  \BibitemOpen
  \bibfield  {author} {\bibinfo {author} {\bibfnamefont {S.}~\bibnamefont
  {Havlin}}, \bibinfo {author} {\bibfnamefont {L.~A.}\ \bibnamefont
  {Braunstein}}, \bibinfo {author} {\bibfnamefont {S.~V.}\ \bibnamefont
  {Buldyrev}}, \bibinfo {author} {\bibfnamefont {R.}~\bibnamefont {Cohen}},
  \bibinfo {author} {\bibfnamefont {T.}~\bibnamefont {Kalisky}}, \bibinfo
  {author} {\bibfnamefont {S.}~\bibnamefont {Sreenivasan}}, \ and\ \bibinfo
  {author} {\bibfnamefont {H.}~\bibnamefont {{Eugene Stanley}}},\ }\href
  {\doibase https://doi.org/10.1016/j.physa.2004.08.053} {\bibfield  {journal}
  {\bibinfo  {journal} {Physica A}\ }\textbf {\bibinfo {volume} {346}},\
  \bibinfo {pages} {82} (\bibinfo {year} {2005})}\BibitemShut {NoStop}%
\bibitem [{\citenamefont {Vaknin}\ \emph {et~al.}(2020)\citenamefont {Vaknin},
  \citenamefont {Gross}, \citenamefont {Buldyrev},\ and\ \citenamefont
  {Havlin}}]{vaknin-20}%
  \BibitemOpen
  \bibfield  {author} {\bibinfo {author} {\bibfnamefont {D.}~\bibnamefont
  {Vaknin}}, \bibinfo {author} {\bibfnamefont {B.}~\bibnamefont {Gross}},
  \bibinfo {author} {\bibfnamefont {S.~V.}\ \bibnamefont {Buldyrev}}, \ and\
  \bibinfo {author} {\bibfnamefont {S.}~\bibnamefont {Havlin}},\ }\href
  {\doibase 10.1103/PhysRevResearch.2.043005} {\bibfield  {journal} {\bibinfo
  {journal} {Phys. Rev. Research}\ }\textbf {\bibinfo {volume} {2}},\ \bibinfo
  {pages} {043005} (\bibinfo {year} {2020})}\BibitemShut {NoStop}%
\bibitem [{\citenamefont {Coniglio}(1982)}]{coni-82}%
  \BibitemOpen
  \bibfield  {author} {\bibinfo {author} {\bibfnamefont {A.}~\bibnamefont
  {Coniglio}},\ }\href {\doibase 10.1088/0305-4470/15/12/032} {\bibfield
  {journal} {\bibinfo  {journal} {Journal of Physics A: Mathematical and
  General}\ }\textbf {\bibinfo {volume} {15}},\ \bibinfo {pages} {3829}
  (\bibinfo {year} {1982})}\BibitemShut {NoStop}%
\bibitem [{\citenamefont {Stauffer}\ and\ \citenamefont
  {Aharony}(1994)}]{stauf-94}%
  \BibitemOpen
  \bibfield  {author} {\bibinfo {author} {\bibfnamefont {D.}~\bibnamefont
  {Stauffer}}\ and\ \bibinfo {author} {\bibfnamefont {A.}~\bibnamefont
  {Aharony}},\ }\href@noop {} {\emph {\bibinfo {title} {Introduction to
  Percolation Theory}}}\ (\bibinfo  {publisher} {Taylor \& Francis},\ \bibinfo
  {year} {1994})\BibitemShut {NoStop}%
\end{thebibliography}
\end{document}